\documentclass[aps,preprint,showpacs,superscriptaddress]{revtex4-1}  
\usepackage{graphicx}
\usepackage{subcaption}
\usepackage{bm}   
\usepackage{color}
\usepackage{amssymb}
\usepackage{amsmath}
\usepackage{float}

\begin{document}
	
	\title[Emergence and mitigation of extreme events]{Emergence and mitigation of extreme events in a parametrically driven system with velocity-dependent potential}
	\author{S. Sudharsan}
	\affiliation{Department of Nonlinear Dynamics, Bharathidasan University, Tiruchirappalli - 620 024, Tamilnadu, India}
	\author{A. Venkatesan}
	\affiliation{Research Department of Physics, Nehru Memorial College (Autonomous), Puthanampatti, Tiruchirappalli - 621 007,  Tamil Nadu, India.}
	\author{P. Muruganandam}
	\affiliation{Department of Physics, Bharathidasan University, Tiruchirappalli 620 024, Tamil Nadu, India.}
	\author{M. Senthilvelan}
	\email[Correspondence to: ]{velan@cnld.bdu.ac.in}
	\affiliation{Department of Nonlinear Dynamics, Bharathidasan University, Tiruchirappalli - 620 024, Tamilnadu, India}
	\vspace{10pt}
	
	\begin{abstract}
		In this paper, we discuss the emergence of extreme events in a parametrically driven non-polynomial mechanical system with a velocity-dependent potential. We confirm the occurrence of extreme events from the probability distribution function of the peaks, which exhibits a long-tail. We also present the mechanism for the occurrence of extreme events. We found that the probability of occurrence of extreme events alternatively increase and decrease with a brief region where the probability is zero. At the point of highest probability of extreme events, when the system is driven externally, we find that the probability decreases to zero. Our investigation confirms that the external drive can be used as an useful tool to mitigate extreme events in this nonlinear dynamical system. Through two parameter diagrams, we also demonstrate the regions where extreme events gets suppressed. In addition to the above, we show that extreme events persits when the sytem is influenced by noise and even gets transformed to super-extreme events when the state variable is influenced by noise.
	\end{abstract}
	
	%
	%
	%
	%
	%
	\maketitle
\section{Introduction}

Extreme events are unusual events observed in dynamical systems. These are classified as rare events because of their tendency to occur less frequently. Their occurrences are found ubiquitously in many natural, engineered, and societal systems such as rogue waves, floods, cyclones, tsunami, tornadoes, earthquakes, droughts, epidemics, epileptic seizures, material ruptures, explosions, nuclear disasters, chemical contamination, financial crisis, share market crashes, and ecological regime shifts \cite{Krause2015,Dysthe2008,Jentsch2005}. The importance of studying these events lies in averting the catastrophic consequences that these events produce. For this, extreme events should be detected in dynamical systems modelling natural, engineered, or societal systems. The ways to mitigate extreme events should also be identified properly.

In the literature, extreme events are identified and reported in a variety of systems ranging from physical, biological, laser, optical fibers, climatic models to electronic circuits, ocean waves, and ecological models~\cite{Dysthe2008,Kumarasamy2018,Ansmann2013,Reinoso2013,Solli2007,Bodai2011,Yukalov2012,Moitra2019,Chaurasia2020}. The occurrence of extreme events is reported in both isolated systems and coupled systems/networks, see for example Refs.~\cite{Kingston2017,Kingston2020,Ray2020,Chowdhury2019}. As far as the dynamical systems are concerned, these events occur in FitzHugh-Nagumo oscillators, Hindmarsh-Rose model, climatic models, Lin\'eard system, memristor-based Lin\'enard system, micromechanical system, electronic circuits, coupled Ikeda map, network of moving agents, network of Josephson junctions, dispersive wave models, Ginzburg-Landau model and nonlinear Schr\"{o}dinger equation~\cite{Kumarasamy2018,Ansmann2013,Bodai2011,Kingston2017,Kingston2020,Ray2020,Chowdhury2019,Karnatak2014,Saha2017,Saha2018,Rings2017,Bialonski2015,Ansmann2016,Mishra2018,Oliveira2016,Ray2019,Cousins2014,Kim2003,Galuzio2014}. Further, in real-time engineering applications, these events are found in laser systems, plasma, optical fibers, and superfluid helium \cite{Reinoso2013,Solli2007,Bailung2011,Ganshin2008}. In addition to the above, the extreme events have also been found to occur in experiments such as epileptic EEG studies in rodents, annular wave flume, and climatic studies \cite{Bodai2011,Pisarchik2018,Toffoli2017}, to name a few.

\color{black} Determining the way in which these extreme events emerge  in the considered system turned out to be the main task in all the above works. To do that, first, one has to detect whether the event that occurred in the considered system is extreme or not. From the dynamical systems point of view, an event is classified as ``extreme event" when the trajectory of $x$ or $\dot{x}$ \color{black} crosses a threshold value very rarely~\cite{Dysthe2008}. This threshold value is assumed to be the sum of mean and $n$ times the standard deviation, where $n$ is greater than or equal to four. Depending on the system under consideration, the choice of $n$ may vary. For example, in the case of the Li\'enard system, $n$ is equal to $8$~ \cite{Kingston2017}, while $n=6$ for the Hindmarsh-Rose model~\cite{Mishra2018}. In the case of Patches of ecological populations, $n=10$ is found to be \color{black} a good choice~\cite{Chaurasia2020}. Whenever extreme events occur, the system's trajectory crosses this threshold value.  This is a preliminary way to detect an extreme event. However, to find out the mechanism behind the origin of such extreme events one has to  make further analysis.

There are several mechanisms reported in the literature for the emergence of extreme events. As far as dynamical systems are concerned, we need to determine the mechanism behind the emergence of extreme events. The mechanism can be found by analysing the bifurcation scenario of the considered system particularly at the point of emergence of extreme events. A majority of the extreme events occur when a chaotic attractor bifurcates into another chaotic attractor or when a periodic orbit bifurcates into a chaotic attractor \cite{Ansmann2013,Kingston2017,Ray2019}. In isolated systems, extreme events are found to occur as a result of an interior crisis, due to intermittent expansion, a stick-slip bifurcation, and collision of a chaotic attractor with the saddle orbit~\cite{Kumarasamy2018,Ansmann2013,Kingston2017,Ray2019}. As far as coupled systems are concerned, extreme events are found to occur by a variety of mechanisms such as the breakdown of quasiperiodicity, instability in the antiphase synchronization~\cite{Mishra2018}, the opening of channel like structure~\cite{Karnatak2014}, and the prototype events acting as precursors~\cite{Ansmann2013}. Besides, they also occur as a result of switching between the librational motion to rotational motion,  the switching between pre-crisis to post-crisis region~\cite{Ray2020}, and at the riddled basin, particularly in the region where both pure and mixed states are present~\cite{Saha2018}. In addition to the above, transient instabilities \cite{Sapsis2018}, attractor bubbling \cite{S.Cavalcante2013} and the influence of noise in multistable systems \cite{Pisarchik2011} are also some of the mechanisms behind the emergence of extreme events. The effect of noise on extreme events has also been studied exhaustively in laser systems \cite{ZamoraMunt2014,ZamoraMunt2013,ZamoraMunt2014a,Ahuja2014}.

The above-mentioned studies concentrate only on the detection of mechanism behind the emergence of extreme events. Several studies have also been made to control the emergence of extreme events. For example, in Li\'enard system time delay feedback has been used to control extreme events \cite{Suresh2018}, in coupled Ikeda map introduction of threshold activated coupling was used as an control measure \cite{Ray2019}, in complex networks, network mobiling is found to control extreme events \cite{Chen2014}, in turbulent flows, closed-loop adaptive control was used \cite{Farazmand2019}, in spatially extended systems, localized perturbations was used \cite{Bialonski2015} and in optically injected lasers, influence of noise suppressed extreme events \cite{ZamoraMunt2014}. Very recently methods of mitigating extreme events have also been identified~\cite{Kaveh2020}. We observed that in most of the studies on the control of extreme events, feedback mechanism only either in the form of noise or time delay has been used to supress it. In Ref. \cite{Farazmand2019}, non-feedback method has been used in the form of damping of modes. The aforementioned mitigation methods are either tough or very expensive to implement. In this work, we report the emergence of extreme events in a parametrically driven, damped mechanical system with velocity-dependent potential and demonstrate the suppression of extreme events upon the introduction of external forcing.

The considered model is a mechanical model~\cite{Nayfeh1979} which is known for its rich nonlinear dynamics~\cite{Venkatesan1997,Venkatesan1998}. Typically, in the absence of parametric forcing, this model represents the dynamics of a particle in a rotating parabola. This mechanical model also describes a motor bike being ridden in a rotating parabolic well in a circus, centrifugation devices, centrifugal filters and industrial hoppers~\cite{Venkatesan1997,Sanderson1977,Bear1984,Lai1997}. Parametrically driven system without external forcing is very interesting because the system is not influenced by any external force but still is being driven inherently, which is more realistic.  Optics is an important area where parametric drive refers to the loss modulation \cite{Kumarasamy2018,Bonatto2017,Metayer2014}. Also, as mentioned earlier, the system that we consider has a velocity dependent potential. Electromagnetic systems under the influence of the Lorentz force such as cyclotrons, mass spectrometers, magnetrons, magnetoplasmadynamic thrusters, and railguns are important applications wherein systems have velocity dependent potential. Extreme events occurring in these systems cause a great destruction not only to human beings but also to the entire ecosystem. So it is essential to detect and control extreme events in these systems. Another dynamical system having a velocity dependent potential, which is of interest, is the oscillator version of pion-pion interaction \cite{DELBOURGO1969}. Extreme events in this system may represent abnormal interactions that may have drastic effects. Thus, our work is a starting point in this direction, and our results will help in redesigning the engineering of these systems so that extreme events can be averted.

So far, in the literature, extreme events were either detected in isolated systems with external forcing or in coupled systems. To the best of our knowledge, this is the first time a parametrically driven system with a velocity dependent potential, without any external forcing, exhibiting extreme events is being reported. Further, extreme events were found to occur in the region of interior crisis in the majority of the existing systems. Here in our work, although there are regions of interior-crisis, we do not find any traces of extreme events there even in the very long run. Surprisingly, we identify extreme events to originate at the point of gradual expansion of the chaotic attractor. Yet another interesting result which we brought out through this study is that the probability of these extreme events is found to increase and decrease alternatively with a period of zero-probability of extreme events. We noticed that the velocity-dependent nature of the potential is the reason behind the generation of extreme events. The most important finding of our work is that the probability of extreme events is found to decrease with the introduction of external forcing and becomes zero for a particular strength of external forcing owing to the increase in the velocity of the system under external forcing. The suppression of extreme events by external forcing turns out to be a very simple way of mitigating extreme events and a way which can be very easily implemented in mechanical systems. We further observe that super-extreme events (events that qualify the threshold qualifier with $n\geq 12$) are produced under the influence of noise \cite{Bonatto2017}. Thus, we have found that extreme events get successfully mitigated in a parametrically driven oscillator with velocity-dependent potential under the influence of external forcing.

This paper is organised in the following way. In Sec.~\ref{sec:2}, we introduce the parametrically driven model and analyze the bifurcation diagram. In Sec.~\ref{sec:3}, we detail the emergence of extreme events. In Sec.~\ref{sec:4}, we investigate the probability of the occurrence of extreme events and the mechanism by which extreme events occur. In Sec.~\ref{sec:5}, we describe the mitigation of extreme events and the mechanism behind it. We study the effect of noise on the extreme events in Sec.~\ref{sec:noise}. Finally, we present our conclusion in Sec.~\ref{sec:6}. 

\section{The Model}
\label{sec:2}

\begin{figure}[!ht]
	\centering
	\includegraphics[width=0.5\textwidth]{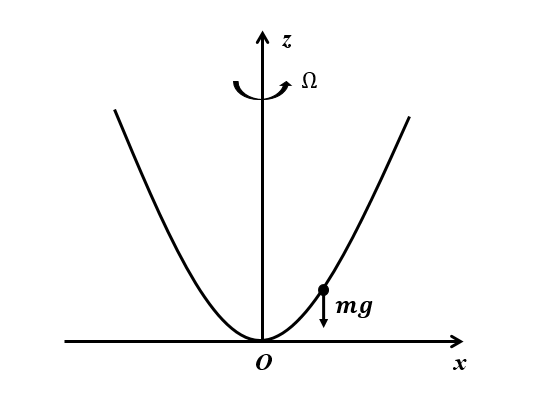}
	\caption{Particle of mass $`m'$ on a rotating parabola defined by $z = \sqrt{\lambda} x^2$, where $g$ is the acceleration due to gravity, $1/\sqrt{\lambda}$ is the semi-latus rectum of the rotating parabola, and $\omega_0$ is the initial angular velocity. and $\Omega$ is the constant angular velocity with which the parabola rotates.}
	\label{system}
\end{figure}

We consider a mechanical model which describes the motion of a freely sliding particle of unit mass on a parabolic wire rotating along the axis of rotation $z = \sqrt{\lambda} x^2$ with a constant angular velocity $\Omega$ ($\Omega^2 = \Omega_0^2 = -\omega_0^2 + g\sqrt{\lambda}$), where $\lambda > 0$ and $\omega_0>0$~\cite{Nayfeh1979} as shown in Fig. \ref{system}. Here $g$ is the acceleration due to gravity, $1/\sqrt{\lambda}$ is the semi-latus rectum of the rotating parabola, and $\omega_0$ is the initial angular velocity. 

The corresponding equation of motion turns out to be

\begin{equation}
(1+\lambda x^2) \ddot{x} + \lambda x \dot{x}^2 + \omega_0^2 x = 0.
\label{first}
\end{equation}
Subjecting Eq.~(\ref{first}) to additional linear damping and periodic forcing, we obtain
\begin{equation}
(1+\lambda x^2) \ddot{x} +\lambda x \dot{x}^2 + \omega_0^2 x + \alpha \dot{x} = f \cos \omega_e t,
\label{master}
\end{equation}
where $\alpha$ is the positive linear damping, while $f$ and $\omega_e$ are the strength and frequency, respectively, of the external force.

In both (\ref{first}) and (\ref{master}) the angular velocity is kept constant. When the angular velocity is considered to be parametrically varying \cite{Nayfeh1979} in the form,
\begin{equation}
\Omega=\Omega_0(1+\epsilon \cos \omega_pt),
\end{equation}
where $\epsilon$ and $\omega_p$ are respectively the strength and frequency of the parametric drive, the equation of motion (\ref{master}) becomes \cite{Nayfeh1979,Venkatesan1997,Venkatesan1998}
\begin{eqnarray}
(1 + \lambda x^2) \ddot{x} + \lambda x \dot{x}^2 + \omega_0^2 x - \Omega_0^2 \left[ 2\epsilon \cos \omega_pt + \frac{1}{2} \epsilon^2 (1 + \cos 2\omega_pt) \right] x + \alpha \dot{x} = f \cos  \omega_e t. 
\label{pardri}
\end{eqnarray}
For numerical integration purpose, we rewrite Eq. (\ref{pardri}) as

\begin{eqnarray}
\dot{x}&=&y,\nonumber\\
\dot{y}&=&\dfrac{f \cos  \omega_e t+\Omega_0^2 \left[ 2\epsilon \cos \omega_pt + \frac{1}{2} \epsilon^2 (1 + \cos 2\omega_pt) \right]x-\alpha y-\lambda x y^2 - \omega_0^2 x}{1 + \lambda x^2}.
\label{couppardri}
\end{eqnarray} 

The numerical integration was carried out using fourth order Runge Kutta method. \color{black} Throughout this work, the parameters are fixed as $\lambda=0.5$, $\omega_0^2 = 0.25$, $\Omega_0^2 = 6.7$, $\omega_p = 1.0$, $\alpha = 0.2$. and $\epsilon$ is the bifurcation parameter. The initial conditions are fixed as $x(0) = 0.1$ and $\dot x (0) = \sqrt{(\epsilon - x(0)^2/4)/(1+x(0)^2/2)}$~ as in \cite{Venkatesan1997} unless otherwise explicitly mentioned. Different routes to strange nonchaotic attractor for the system (\ref{pardri}) were reported in  \cite{Venkatesan1997,Venkatesan1998}. But no studies have been made in the context of extreme events in this system so far. In the case of isolated systems, extreme events are found to occur inside a chaotic orbit. Thus extreme events may occur as a result of interior crisis or during the transition from periodic orbit to chaotic attractor or during the transition from chaos 1 to chaos 2. In the present work, we intend to examine the way extreme events occur in (\ref{pardri}). To make our study systematic, first we analyse the emergence of extreme events in the absence of external forcing, that is $f=0$.

\begin{figure}[!ht]
	\centering
	\includegraphics[width=0.5\textwidth]{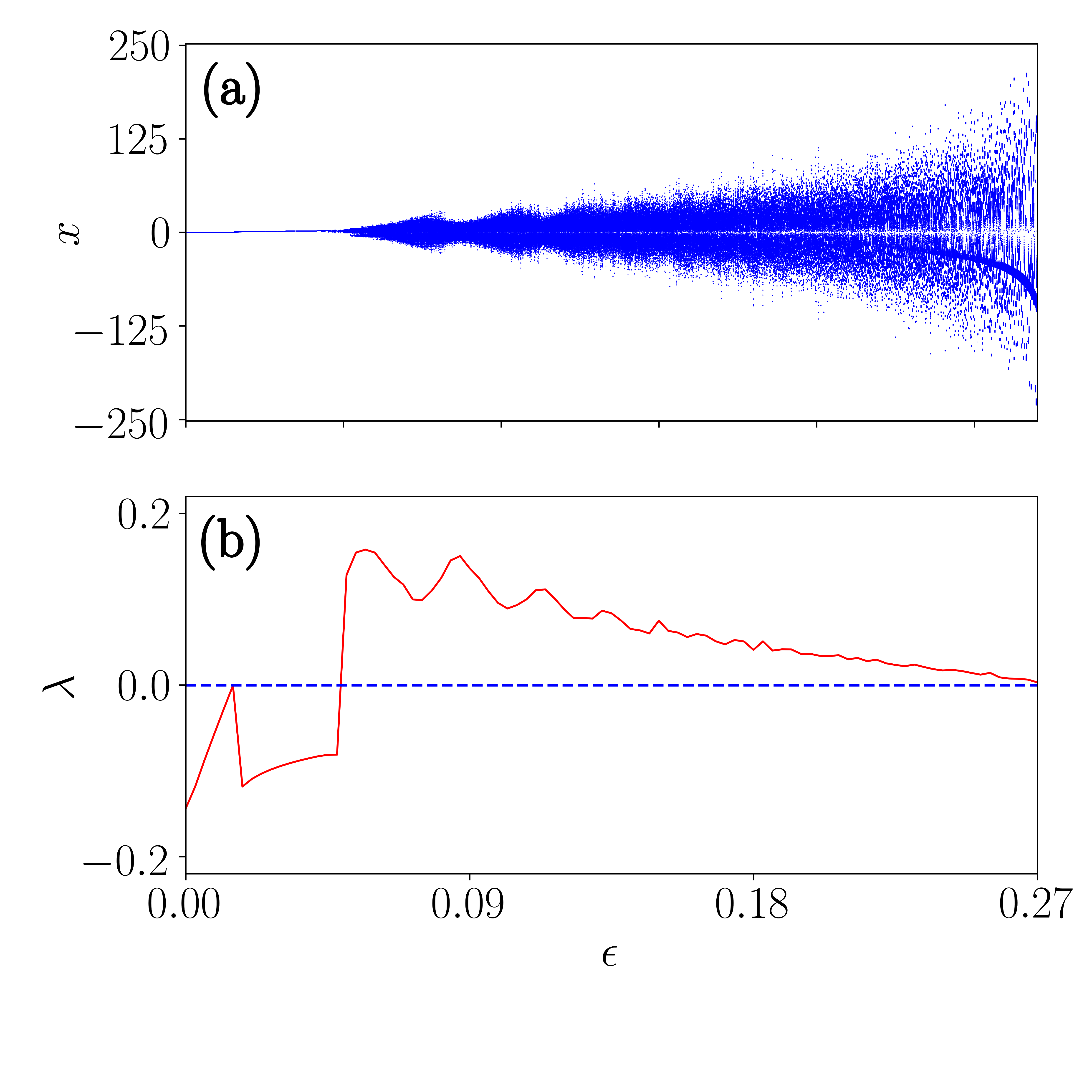}
	\caption{$(a)$~Bifurcation plot and $(b)$ Lypunov Exponent of Eq. (\ref{pardri}). The corresponding parameter values are $\lambda=0.5$, $\omega_0^2=0.25$, $\Omega_0^2=6.7$, $\omega_p=1.0$, $\alpha=0.2$, $f=0$, and $\omega_e=0$.}
	\label{figbif}
\end{figure}

To analyse the emergence of extreme events, to begin, we study the dynamics of system (\ref{pardri}). For this, we plot the bifurcation diagram in Fig.~\ref{figbif}~(a) by varying the parametric drive strength $\epsilon$ in the range $\epsilon=(0,2.7)$. The bifurcation diagram is generated by numerically solving Eq.~(\ref{pardri}) and collecting all the peak values of $x$ . It is clear from the bifurcation diagram that the system transits from regular attractor to chaotic attractor through period doubling route. It is also interesting to note that in the chaotic region there is an alternate expansion and contraction in the chaotic attractor. The corresponding Lyapunov exponent $\lambda$ \cite{Wolf1985} for the system (\ref{pardri}) is shown in Fig.~\ref{figbif}(b). We can visualize a change of sign in the value of Lyapunov exponent from negative to positive confirming the emergence of chaos at $\epsilon=0.05$. Along with the expansion and contraction of the chaotic attractor, Lyapunov exponent also decreases and increases respectively. After every expansion, the size of contracting region decreases and at a particular region only continuous expansion prevails.  Correspondingly, we can observe the Lyapunov exponent to decrease gradually to zero at $\epsilon=2.7$. The decrease in the value of Lyapunov exponent represents the weakening of the chaotic attractor. Finally when the Lyapunov exponent is zero, the system becomes quasi-periodic. 

\section{Observation of Extreme Events}
\label{sec:3}

An event is said to be extreme if the system's trajectory traverses a threshold value $x_{ee}$ \cite{Dysthe2008}.  This threshold value is calculated using the formula $x_{ee}=\langle x \rangle + n\sigma_x$. Here, $\langle x \rangle$ is the mean peak amplitude and $\sigma_x$ is the standard deviation. The mean peak amplitude is the mean of all $max\{x\}$ collected for a very long time interval of about $10^9$ iterations. The standard deviation is calculated through the formula $\sigma_x=\sqrt{\dfrac{(x_i-\langle x \rangle)^2}{\text{Total Number of Peaks}}}$. Finally, the threshold is calculated using the formula $x_{ee}=\langle x \rangle + n\sigma_x$. Here $n \ge 4$ \cite{Ray2019}. This threshold value is called qualifier threshold. The crossing of the system's trajectory over the qualifier threshold $x_{ee}$ can happen even at a very large time. Throughout this work we fix $n=4$ and calculate the value of $x_{ee}$ from very long time iterations, typically of the order of $10\,000\,000$ time units. 

\begin{figure}[!ht]
	\centering
	\includegraphics[width=1.0\textwidth]{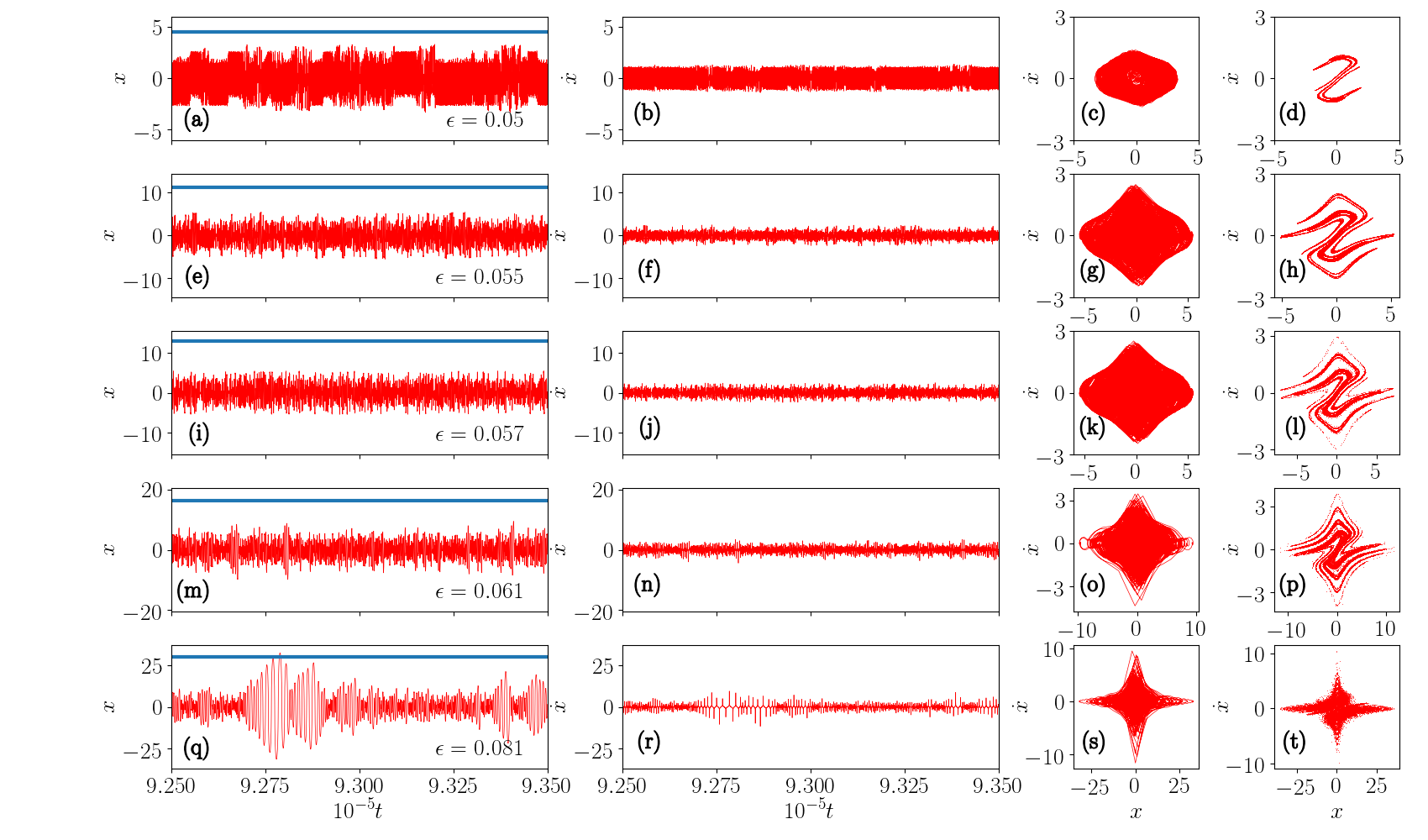}
	\caption{The $x$ series (first panel), $\dot{x}$ series (second panel), the corresponding phase portraits (third panel) and Poincar\'e cross section of system (\ref{pardri}) for $\epsilon=0.05$, $0.055$, $0.057$, $0.061$, and $0.081$. The horizontal line in the first panel correspond to the calculated threshold value $x_{ee}$. The value of $\dot{x}$ is calculated using Eq. (\ref{couppardri}) The other parameters are the same as in Fig.~\ref{figbif}.}
	\label{fig2}
\end{figure}

In Fig.~\ref{fig2}, we plot the time series of $x$ (panel 1), $\dot{x}$ (panel 2), phase portrait (panel 3) and the corresponding Poincar\'e cross section (panel 4). We present these plots for $\epsilon=0.05$, $0.055$, $0.057$, $0.061$, and $0.081$ in order to show how the system (\ref{pardri}) gradually transits from bounded chaotic motion to chaotic motion with large amplitude oscillation or spikes or bursts and where extreme events are produced. The horizontal blue line in panel 1 of Fig. \ref{fig2} represents the threshold value.

System (\ref{pardri}) exhibits chaos for $\epsilon \ge 0.05$. Hence when we increase $\epsilon$ above $0.05$, a sudden expansion of chaotic attractor occurs as a result of an interior crisis. Such interior crisis occur at several places initially. Afterwards, the chaotic attractor expands and shrinks continuously with an overall trend of expansion, as illustrated in Fig.~\ref{figbif}(a). From the origin of chaotic attractor at $\epsilon=0.05$, and until $\epsilon = 0.078$ we do not see any signature of extreme events. The initiation of extreme events occurs only at $\epsilon=0.079$.

At $\epsilon=0.05$, the system exhibits bounded chaos with small amplitude chaotic oscillation as shown in Figs.~\ref{fig2}(a) and \ref{fig2}(c). This bounded chaotic nature is reflected as dense region in the Poincar\'e cross section shown in Fig.~\ref{fig2}(d). Here dense region represents the dense mid region in Fig.~\ref{fig2}(d) without any large amplitude oscillation or spikes or burst. As the parameter $\epsilon$ is further increased, chaotic expansion occurs. At $\epsilon=0.055$, after the chaotic expansion, we observe system's dynamics to have relatively higher number of spikes. While comparing the time series which can be seen in Figs.~\ref{fig2}(a-b) and \ref{fig2}(e-f), phase portraits in Figs.~\ref{fig2}(c) and \ref{fig2}(g) and Poincar\'e cross sections in Figs.~\ref{fig2}(d) and \ref{fig2}(h), we infer that the time series (both $x$ and $\dot{x}$) which can be seen in Figs.~\ref{fig2}(e) and (f) have more number of spikes when compared to the time series in Figs.~\ref{fig2}(a) and (b). We can visualize the effect of increase in the number of spikes through the increase in the size of attractor in the phase portrait in Fig.~\ref{fig2}(g) and through the extended region in the Poincar\'e cross section in Fig.~\ref{fig2}(h). As one visualizes, there are no sparser points in the extended region. When $\epsilon=0.057$, we notice further increase in the spikes in both the time series of $x$ and $\dot{x}$ as shown in Figs. \ref{fig2}(i) and (j). Slight increase in the size of chaotic attractor can be seen in Fig. \ref{fig2}(k) and as a result we observe sparse points surrounding the previously extended region. A similar increase in the number of spikes is also observed for $\epsilon=0.061$ and is presented in the Figs.~\ref{fig2}(m-p). Until $\epsilon=0.078$, we do not observe any extreme events and we notice that the threshold (horizontal line in the time series plots) is well above and no trajectory crosses it. We observe that extreme events starts to originate at $\epsilon=0.079$. In particular, when $\epsilon=0.081$, from Fig.~\ref{fig2}(q), we see that the trajectory crosses the threshold once and this corresponds to an extreme event.  When $\epsilon=0.081$, there are considerably more number of spikes and bursts (see Fig. \ref{fig2}(q) and (r)), significant increase in the size of phase portrait and extended regions in the Poincar\'e cross section (see Figs. \ref{fig2}(s) and (t)). From these observations, we conclude that as the value of $\epsilon$ increases, the number of extra spikes present in the time series also increases. Correspondingly, the size of the chaotic attractor increases in the phase portrait for every increase of $\epsilon$. As far as the Poincar\'e cross section is concerned, we can see the extension of mid region and the presence of sparser points every time when $\epsilon$ is increased. This is similar to what was observed in \cite{Ray2019}. For all the values of $\epsilon$, although the nature of $\dot{x}$ is qualitatively similar to $x$ in producing additional spikes after every increase in $\epsilon$, we find $\dot{x}$ to be low in amplitude throughout the time when compared with the $x$ - time series  
\color{black}

\begin{figure}[!ht]
	\centering
	\includegraphics[width=1.0\textwidth]{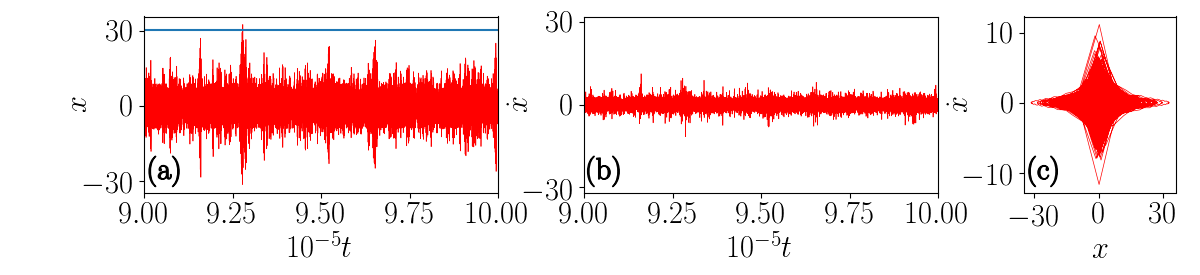}
	\caption{Plots showing (a) the time series of $x$ and (b) the time series of $\dot{x}$ and (c) the corresponding phase portrait for a larger time domain at $\epsilon=0.081$. The horizontal line in (a) represents the threshold value. The value of $\dot{x}$ is calculated using Eq. (\ref{couppardri}) The other parameters are the same as in Fig.~\ref{figbif}.}
	\label{fig3}
\end{figure}

Further, to verify whether this extreme event is a lone event, we plot the time series in both $x$ and $\dot{x}$ and the corresponding phase portrait for nearly $100\, 000$ time units in Fig.~\ref{fig3}(a-c). We find that only one event occurs in $100\, 000$ time units. The Poincar\'e cross section will be the same as Fig. (\ref{fig2})(t). %
In order to know precisely the probability of different value of peaks that crosses the threshold, we plot the probability distribution function (PDF) of peaks ($P_n$) in Fig.~\ref{pdf}. Extreme events are confirmed by the fat-tailed distribution. Since there is a finite probability for the occurrence of peaks beyond the threshold value, we obtain such a fat-tailed distribution.

\section{Occurrence Probability and Mechanism}
\label{sec:4}

As mentioned earlier, the size of the chaotic attractor alternatively expands and shrinks continuously after the third interior crisis. The region of expansion is more when compared to that of shrinking. One surprising fact which we observe is that in the course of expansion and shrinkage of chaotic attractor, the occurrence of extreme events also increases and decreases, respectively. Metayer et al. observed a similar behaviour, where an increase and decrease in the probability of extreme events that takes place inside a chaotic attractor prevails to be constant in size until the next crisis (expansion) ~\cite{Metayer2014}. But, in the present case, the increase and decrease in the probability of extreme events does not emerge in the regions of crisis but occurs along the expansion and contraction of the chaotic attractor.%
\begin{figure}[!ht]
	\centering
	\includegraphics[width=0.5\textwidth]{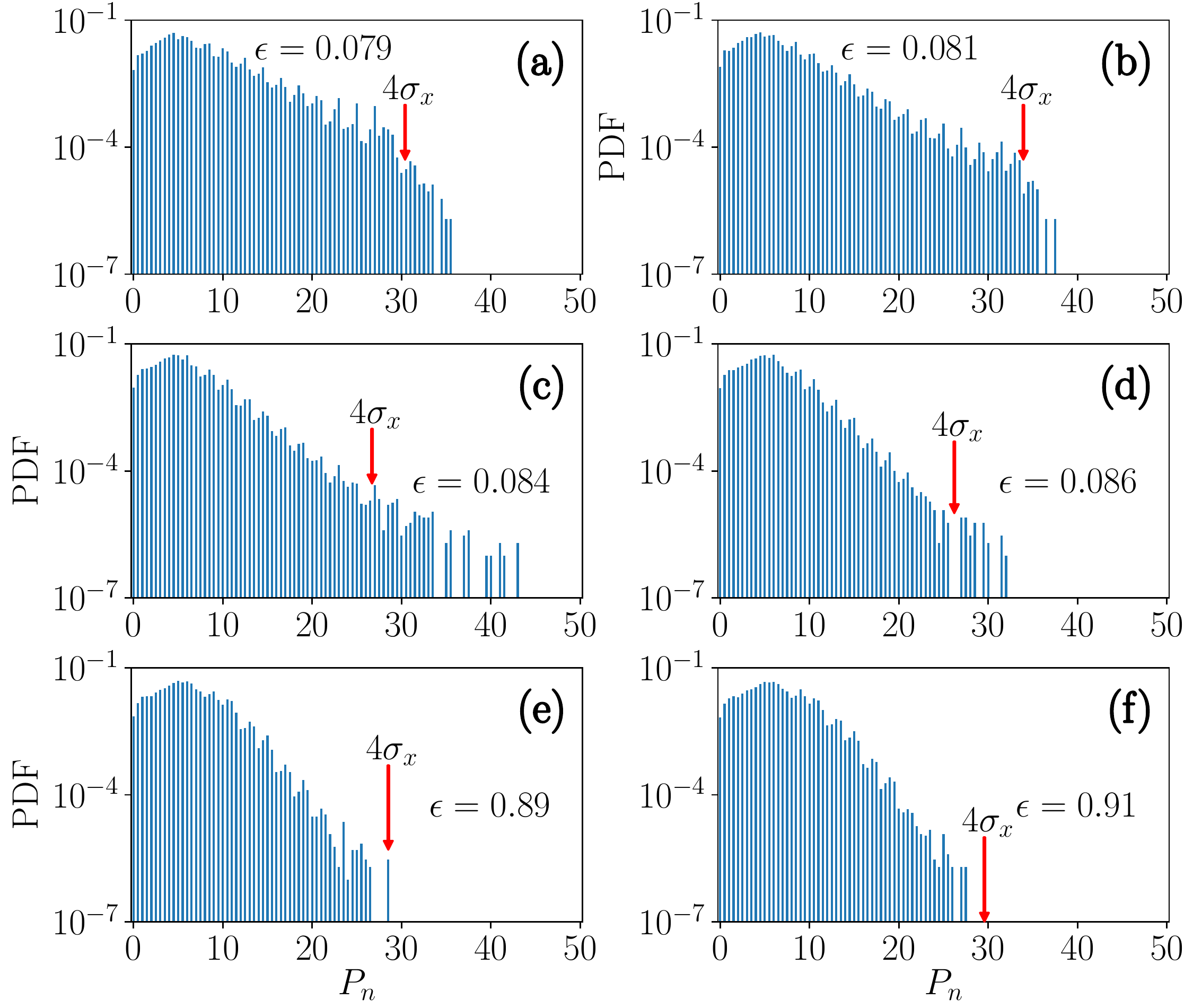}
	\caption{Plots of the probability distribution function (PDF) of peaks ($P_n$) for system (\ref{pardri}) with (a) $\epsilon=0.079$,  (b) $\epsilon=0.081$, (c)  $\epsilon=0.084$, (d) $\epsilon=0.086$, (d) $\epsilon=0.089$ and (f) $\epsilon=0.091$. The qualifier threshold $x_{ee}=\langle x \rangle + 4\sigma_x$ is noted by red vertical arrows and represented just as $4\sigma_x$ (for the purpose of convenience). The other parameters are the same as in Fig.~\ref{fig3}. The qualifier threshold $x_{ee}=\langle x \rangle + 4\sigma_x$ is noted by red vertical arrows and represented just as $4\sigma_x$ (for the purpose of convenience)}
	\label{pdf}
\end{figure}%
~In the following, we produce proper evidences in the form of peak PDF and probability plot for emphasizing the fact that probability of extreme events emerge and then alternatively increase and decrease along with the expansion and contraction of the chaotic attractor. Since peak PDF's confirm whether or not the extreme event has occurred, we plot the peak PDF's in Figs.~\ref{pdf}(a) - \ref{pdf}(f) respectively for $\epsilon=0.079$,  $0.081$, $0.084$, $0.086$, $0.089$ and $0.091$ along with the contraction and expansion of chaotic attractor and infer how the changes occur.  In all the cases the qualifier threshold $x_{ee}=\langle x \rangle + 4\sigma_x$ is noted by red vertical arrows and represented just as $4\sigma_x$ (for the purpose of convenience).

In Fig.~\ref{pdf}(a), we plot the PDF of peaks when $\epsilon=0.079$, and we can notice that only few peaks are present beyond the threshold and the probability is also comparitively low. \color{black} It is the point where extreme events get initiated. The range of peak values beyond the threshold gets increased step by step until a point and then decreases to zero systematically. In Fig.~\ref{pdf}(c), when $\epsilon=0.084$, we find the maximum range of values of peaks occupying beyond the threshold. Afterward, increasing $\epsilon$ further decreases the range of peaks until $\epsilon=0.089$. At $\epsilon=0.091$ there are no occurrences of extreme events. This fact can be seen in Figs. \ref{pdf}(e) and \ref{pdf}(f). By and large, in Fig.~\ref{pdf}(e), only a relatively small range of peak values are found beyond the threshold. As far as Fig.~\ref{pdf} is concerned, we can only tell whether or not the extreme events occur and the range of peak values that are present beyond the threshold. 

\begin{figure}[!ht]
	\centering
	\includegraphics[width=0.5\textwidth]{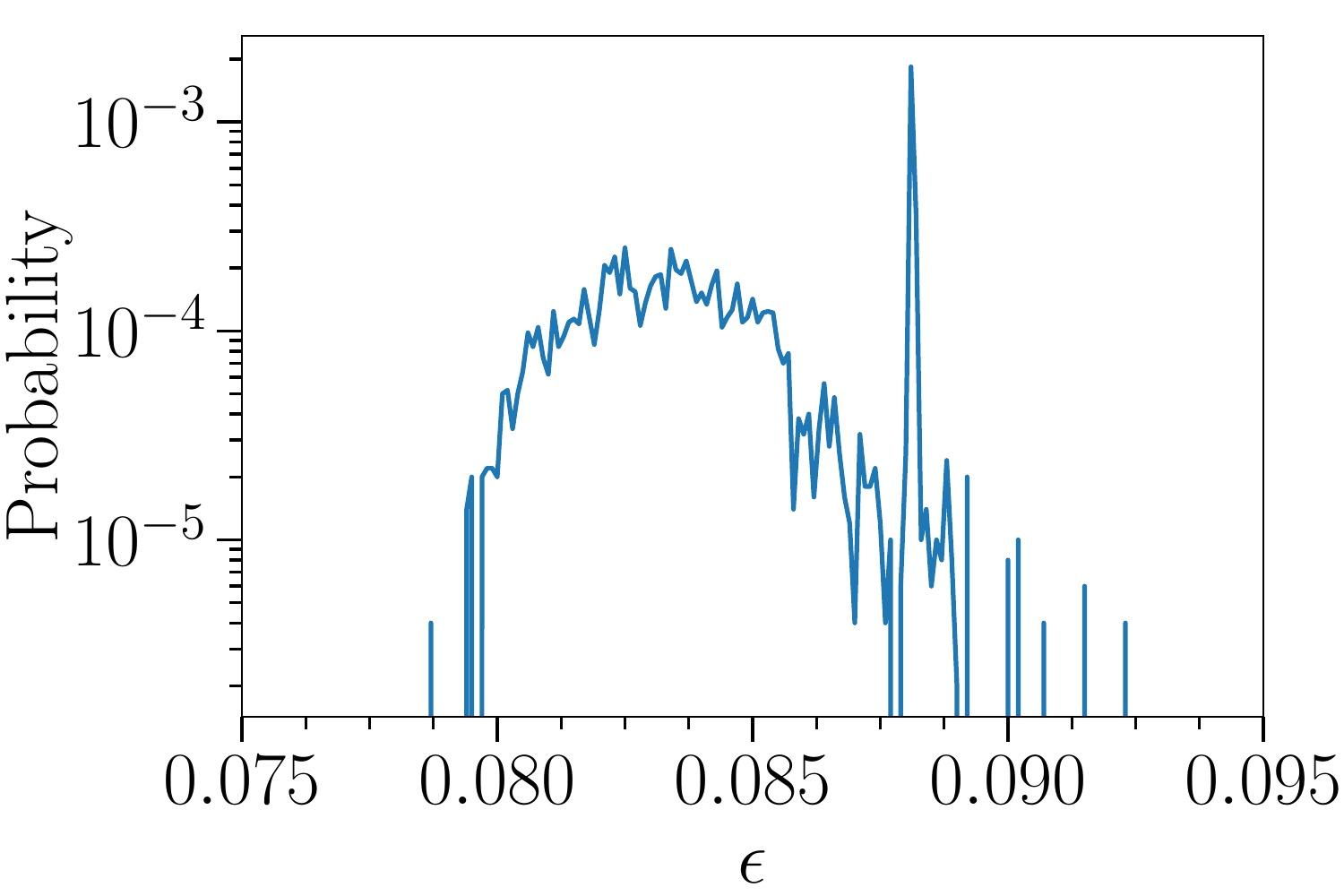}
	\caption{Plot of the probability vs $\epsilon$ showing the variation of probability of the occurrence of extreme events for the range of $\epsilon$ from $0.075$ to $0.095$. It is explicit that initiation of extreme events occur at $\epsilon=0.079$ and terminates at $\epsilon=0.090$.}
	\label{fig6}
\end{figure}

To know exactly the value of probability at which extreme events occur, we calculate the probability of the occurrence of extreme events and plot it against $\epsilon$ in Fig.~\ref{fig6}. From Fig.~\ref{fig6}, we can infer that the probability for the occurrence of extreme event is zero until $\epsilon=0.078$. Only at $\epsilon=0.079$, a non-zero probability for the occurrence of extreme event occurs confirming the emergence of extreme events. This probability increases gradually until $\epsilon=0.081$ and decreases afterward and becomes completely zero at $\epsilon=0.090$. %
Eventhough there are a few points beyond $\epsilon=0.090$ where the probability is non-zero, it occurs discontinuously but becomes zero afterward. Note that even though the range of peak values is large at $\epsilon=0.084$, the probability of the occurrence of extreme events is large only for $\epsilon=0.081$. In Fig. (\ref{fig6}) we note that when the probability of the occurrence of extreme events is decreasing with varying $\epsilon$, suddenly, there is a huge jump in the probability of the system at $\epsilon=0.0881$. %
\begin{figure}[!ht]
	\centering
	\includegraphics[width=0.5\textwidth]{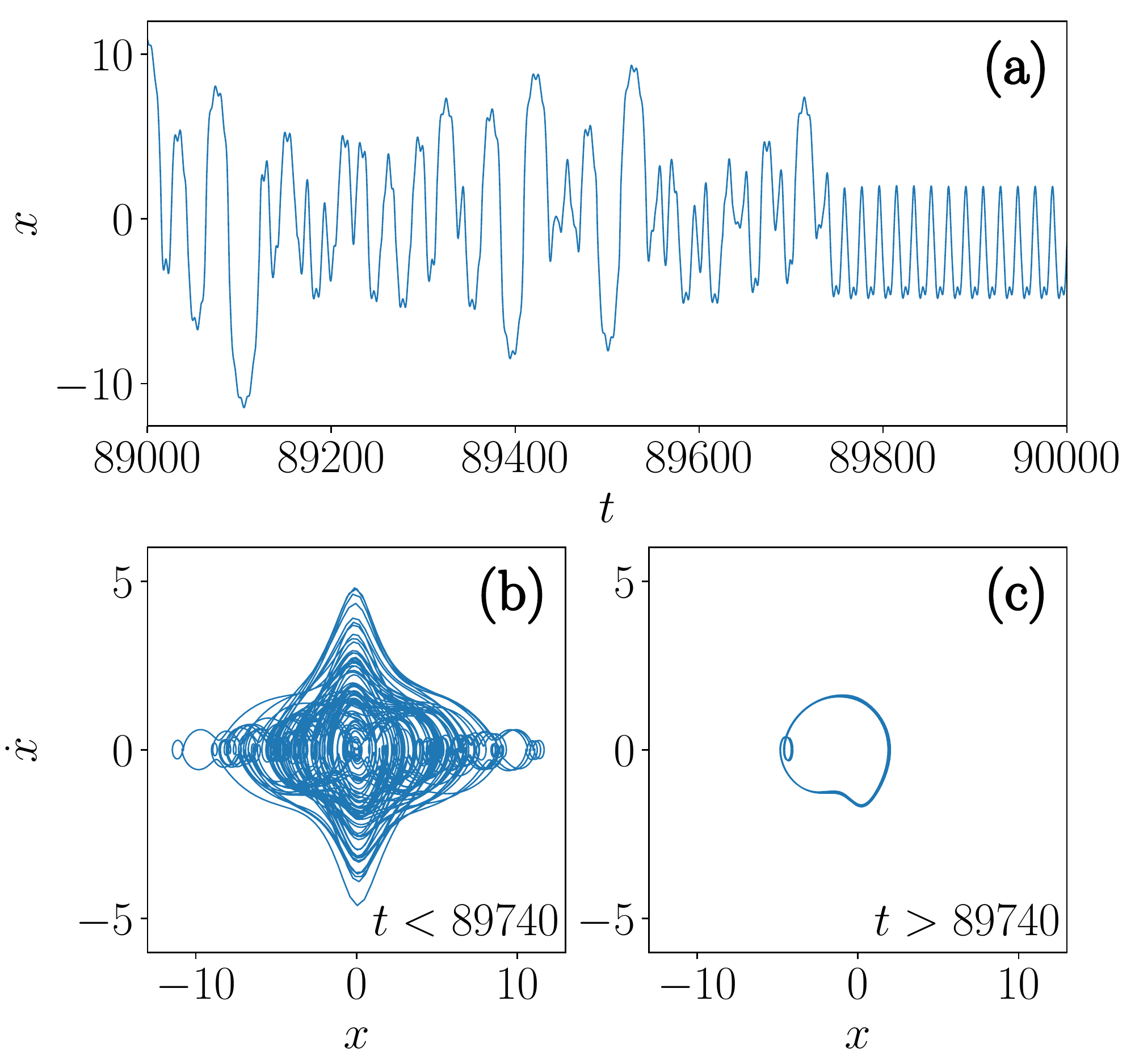}
	\caption{(a) Plot of time series showing the long transient behaviour at $\epsilon=0.0881$ corresponding to the sudden increase in peak as in Fig. (\ref{fig6}). Phase portraits of (a) transient chaotic attractor for $t < 89740$ and (b) the final period-2 orbit for $t > 89740$. The initial condition is chosen as $x(0) = 0.2$ and $\dot x(0) \approx 0.2767$ [$\dot x (0) = \sqrt{(\epsilon - x(0)^2/4)/(1+x(0)^2/2)}$].}
	\label{fig6-ts}
\end{figure}%
It occurs at that particular value of $\epsilon$ alone due to the long transient behaviour of the system shown in Fig. (\ref{fig6-ts}). The system (\ref{pardri}) is exhibiting a very long chaotic transient, and after that, it displays period-2 oscillations with relatively small amplitudes. The transition from chaos to periodic behaviour can be clearly seen from the time series in Fig.~\ref{fig6-ts}(a). At $t=89~740$, this transition occurs. The phase portrait of the chaotic nature before the transient and the phase portrait of the periodic nature after the transient are shown in Figs.~\ref{fig6-ts}(b) and \ref{fig6-ts}(c) respectively. The contribution of these small-amplitude periodic oscillations reduces the threshold and thereby increasing the probability drastically. 
This periodic behaviour after a long chaotic transient, at a narrow region of $\epsilon$ close to $0.0881$, is the reason behind the sharp spike in the probability as seen in Fig. \ref{fig6}. One may note that at this particular value of $\epsilon$, transient time varies depending on the choice of the initial conditions, but still it exhibits long transient behaviour. When a system exhibits similar long transient behaviour, due to periodic nature beyond the long transient time, the value of the threshold will be lowered which makes almost all large peaks in the transient to cross the threshold which contradicts the basic definition of extreme events. One main inference what we make from this is, if suppose a system exhibits long transient behaviour, then one should make the calculation of threshold only after letting out the transients.  Further, to have a complete overview of the probability of the occurrence of extreme events for the entire chaotic region, we compare the bifurcation diagram with the probability and $d_{max}$ plot in Fig.~\ref{prob_epilon_full}. Before going into the details, in the following, we discuss briefly about the precursor behind the generation of extreme event in the system (\ref{pardri}).


From Fig.~\ref{fig3}(c), we notice that the system exhibits dynamics in such a way that the trajectory of $x$ excurses to maximum whenever $\dot{x}$ is zero and vice-versa (similar to simple harmonic oscillator). This shows that whenever the velocity of the system is low, the trajectory of the system travels a longer distance in phase space, making it larger than the normal. When the velocity is high, $x$ travels a least distance. Thus, we can conclude that in (\ref{pardri}), the velocity of the system acts as a precursor in producing extreme events. 

To analyze the probability of the occurrence of extreme events in the entire chaotic regime we plot Fig.~\ref{prob_epilon_full}. %
\begin{figure}[!ht]
	\centering
	\includegraphics[width=0.5\textwidth]{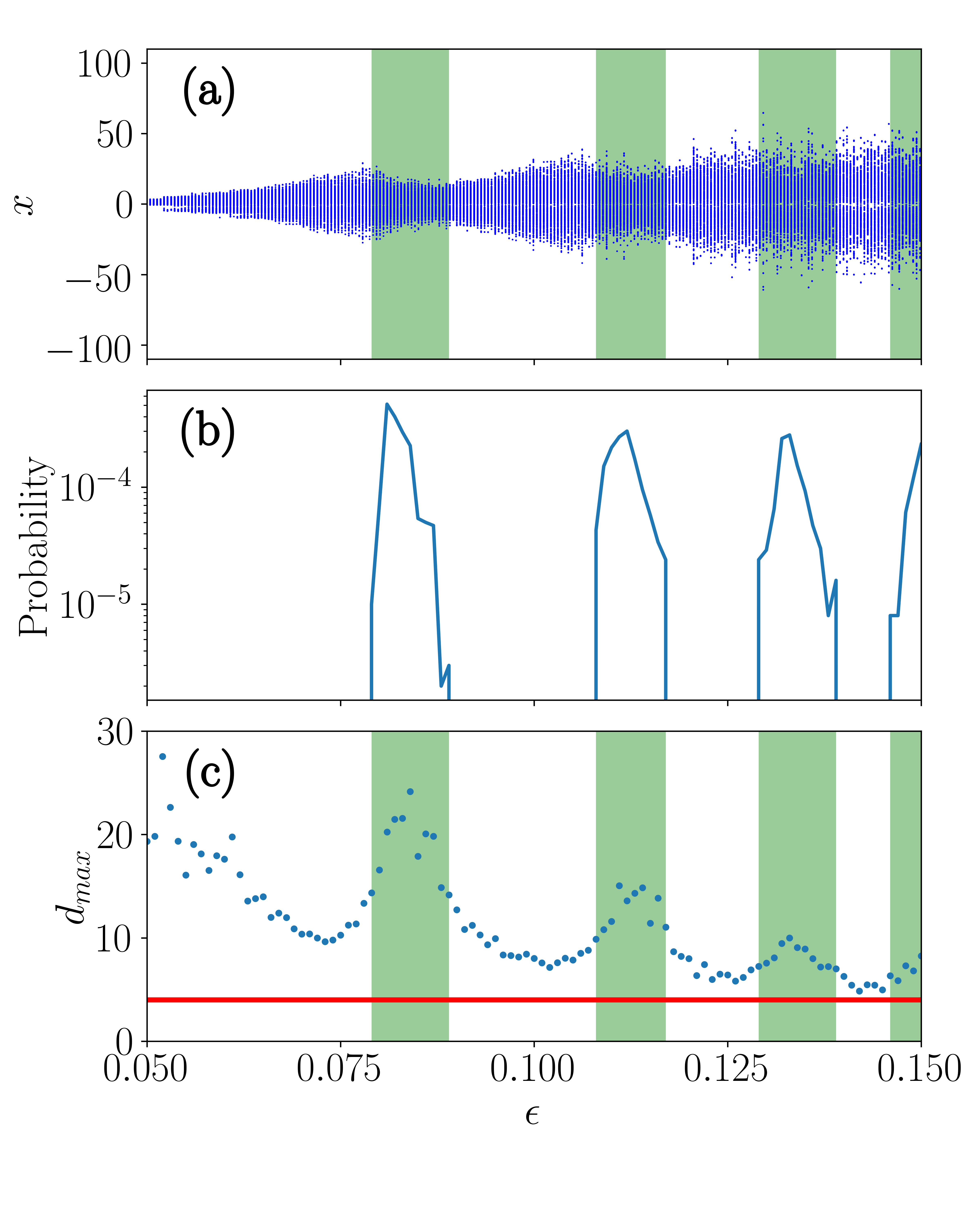}
	\caption{Plots of (a) bifurcation diagram same as \ref{figbif} with additional zoning, (b) the corresponding probability of the occurrence of extreme events  and (c) $d_{max}$ against the parameter $\epsilon$. The horizontal red line in (c) represents the $n=4$ qualifier. The shaded areas in (a) \& (c) represents the regions with non zero probability for the occurrence extreme events for the corresponding $\epsilon$ values.}
	\label{prob_epilon_full}
\end{figure}%
It can be seen that the probability of the occurrence of extreme events initiates at a point, increases and reaches a maximum point, decreases and becomes zero at a particular point. Once the probability becomes zero thereafter the probability remains zero until the next emergence of extreme events. During the next emergence also, extreme events initiate at a particular point and repeats the same steps as in the previous region. We had already mentioned that whenever the velocity of the system is maximum, $x$ takes the least value. So, the velocity of the system increases as soon as the contraction phase of the chaotic attractor begins. This decrease in the velocity of the system, tries to pull down the $x$ trajectory from maximum. In this connection, amplitude of many peaks decreases and only few peaks in the entire time domain have very large amplitude and it crosses the threshold. The regions of non-zero probability can be identified by the shaded regions in Fig.~\ref{prob_epilon_full}(a). Extreme events occur only in these regions. We find four such regions in the bifurcation diagram where the extreme events occur. Also, it is obvious from Fig.~\ref{prob_epilon_full} that extreme events initiate in the vicinity of the larger peaks and has a high probability of occurrence at the maximum size of the attractor. The probability decreases gradually along with the size of the attractor and becomes almost zero near to the point where the continuous shrinking of chaotic attractor stops and a continuous expansion of chaotic attractor again begins. This continues again and again until a stage is reached where the attractor exponentially grows. Also the size of continuous expansion decreases after every shrinking thus reaching a point where only expansion takes place. This is also clear from Fig.~\ref{prob_epilon_full}(b) where the distance between the point at which the extreme events become zero and the point at which the extreme events initiate decreases. In Fig.~\ref{prob_epilon_full}(c), we have shown the variation of $d_{max}$ against $\epsilon$. the value of $d_{max}$ is calculated using the formula \cite{Ray2019}. 
\begin{equation}
d_{max}=\dfrac{\mathrm{max}(x_i)-\langle x\rangle}{\sigma_x}.
\end{equation}
Observing Fig.~\ref{prob_epilon_full}(c), we find that the value of $d_{max}$ decreases with expanding attractor and starts increasing with contracting attractor. As we have already mentioned, extreme events initiate when the chaotic attractor starts to decrease in size. So the value of $d_{max}$ raises whenever the extreme occurs, reaches a peak value and decreases thereafter until next extreme event occurs. Although there is an increase and decrease in the value of $d_{max}$ inside the region of occurrence of extreme events, the lowest value inside the region of occurrence of extreme events is still higher than the previous lowest value outside the region of occurrence of extreme events. The increase in the value of $d_{max}$ during the emergence of extreme events confirms the point of emergence of extreme events.  

\begin{figure}[!ht]
	\centering
	\includegraphics[width=1.0\textwidth]{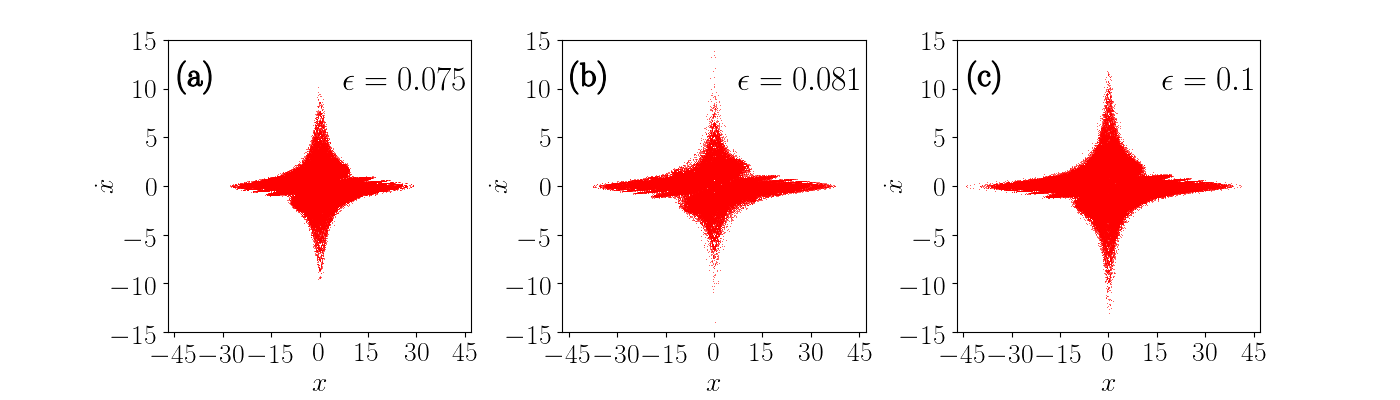}
	\caption{Poincar\'e section at the points inside the region of non-extreme events (a) \& (c) and at the point inside the region of extreme events (b)}
	\label{poincare}
\end{figure}

Further to see the difference between the dynamics between the regions of extreme events and regions of non-extreme events, we plot the Poincar\'e cross section at three different points in Fig.~\ref{poincare}. In particular, Figs.~\ref{poincare}(a) and (c) represent point from non-extreme event regions and Fig.~\ref{poincare}(b) represents the point from the region of extreme event.

Although there is not much difference between the three figures, still the Poincar\'e cross section at the point inside the region of extreme event shown in Fig.~\ref{poincare}(b) is sparser than the Figs.~\ref{poincare}(a) and (c). 
\color{black}

\section{Mitigation}
\label{sec:5}

In Sec.~\ref{sec:4}, we focussed our attention on how extreme events occur in the system (\ref{pardri}) without any external forcing. When we intended to analyse the effect of external forcing on system (\ref{pardri}), we find that external forcing acts as a simple tool in mitigating extreme events in (\ref{pardri}).  One of the salient features of the study of extreme events is to investigate how extreme events can be controlled in the dynamical system under concern. With this aim, now we inspect in what way the extreme events in the system (\ref{pardri}) can be suppressed while externally driving the system by a periodic force.%

\begin{figure}[!ht]
	\centering
	\includegraphics[width=0.75\textwidth]{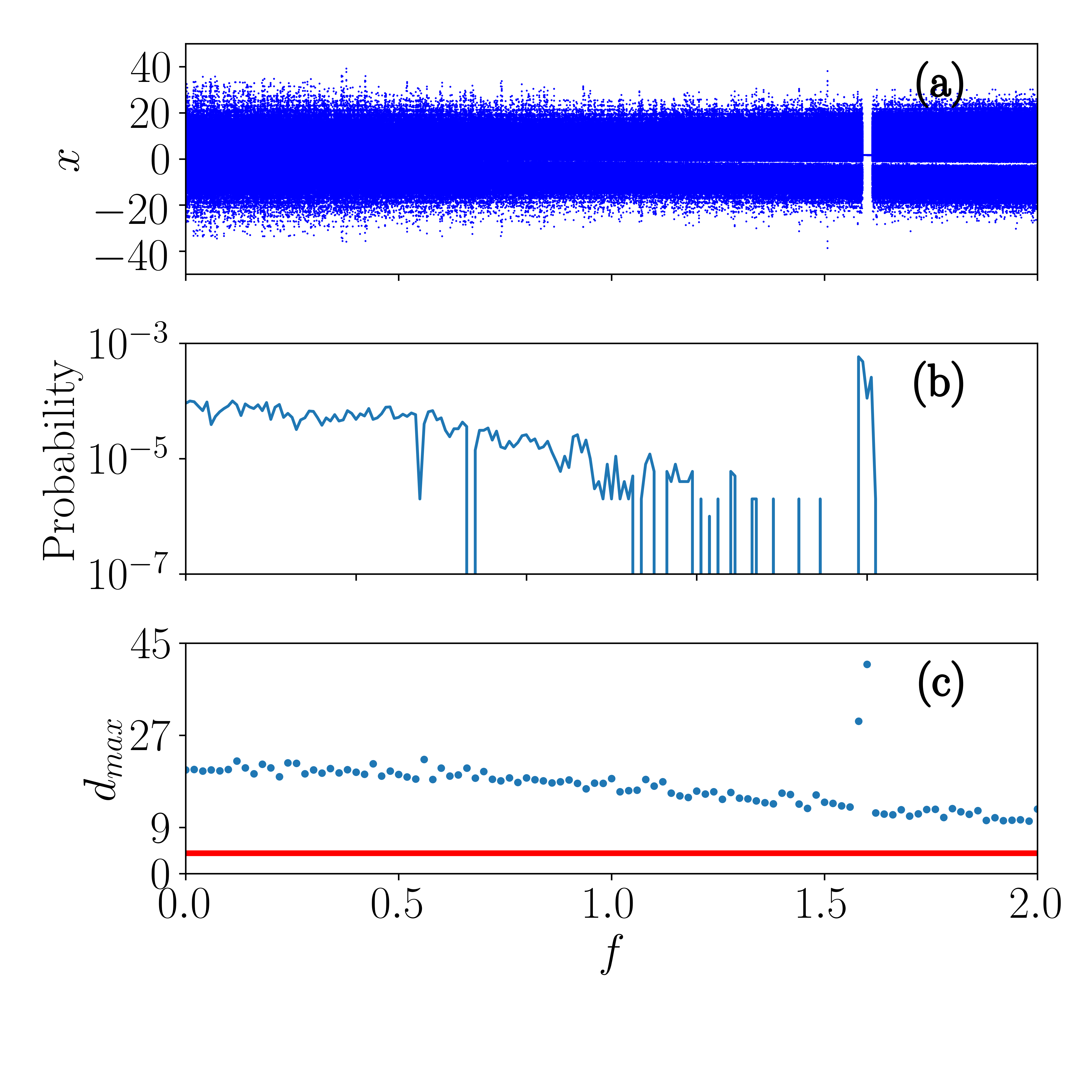}
	\caption{Plots of (a) bifurcation diagram (b) variation of the probability of the occurrence of extreme events and (c) $d_{max}$ plot as a function of the external forcing amplitude $f$. The horizontal red line in (c) represents the $n=4$ qualifier. The other parameters are fixed as $\lambda=0.5$, $\omega_0^2=0.25$, $\Omega_0^2=6.7$, $\omega_p=1.0$, $\alpha=0.2$, $\epsilon = 0.081$ and $\omega_e = 1$.}
	\label{sup}
\end{figure}%

For this, we introduce an external force of the form  $f \cos \omega_e t$ to the system (\ref{pardri}) with frequency $\omega_e=1.0$, and vary the amplitude $f$. For this purpose, we fix the value of $\epsilon$ as $0.081$. In Fig.~\ref{sup} we show the variation of the probability of the occurrence of extreme events as a function of the external forcing amplitude $f$.

On increasing the strength of the external force $f$ above zero in the range $(0,2)$, we noticed a gradual decrease in the probability of the occurrence of extreme events as shown in Fig.~\ref{sup}(b). We also observe at least finite probability of the occurrence of extreme events for $f<1.6$, and zero probability for $f\geq1.7$. The corresponding bifurcation diagram is shown in Fig.~\ref{sup}(a). We can see the gradual decrease in the sparser points as $f$ varies. But still the large amplitude oscillation persists throughout the parameter space. This confirms that the chaotic nature prevails even for large $f$. Further, the $d_{max}$ plot which is given in Fig.~\ref{sup}(c) shows that $d_{max}$ decreases with increasing $f$. This confirms the vanishing of sparser points as $f$ increases. This kind of suppression is an intriguing feature that occur in (\ref{pardri}). Usually, in many single systems, extreme events occur on the introduction of external forcing. But surprisingly, extreme events get suppressed in (\ref{pardri}) upon introducing the external forcing. It is an unexpected phenomenon that occur in (\ref{pardri}), which is a very useful tool for mitigation purposes. The sudden abnormal increase in the probability in Fig.~\ref{sup} at $f=1.6$ can be attributed to the long transient behaviour as observed in Fig.~\ref{fig6}. %
\begin{figure}[!ht]
	\centering
	\includegraphics[width=0.5\textwidth]{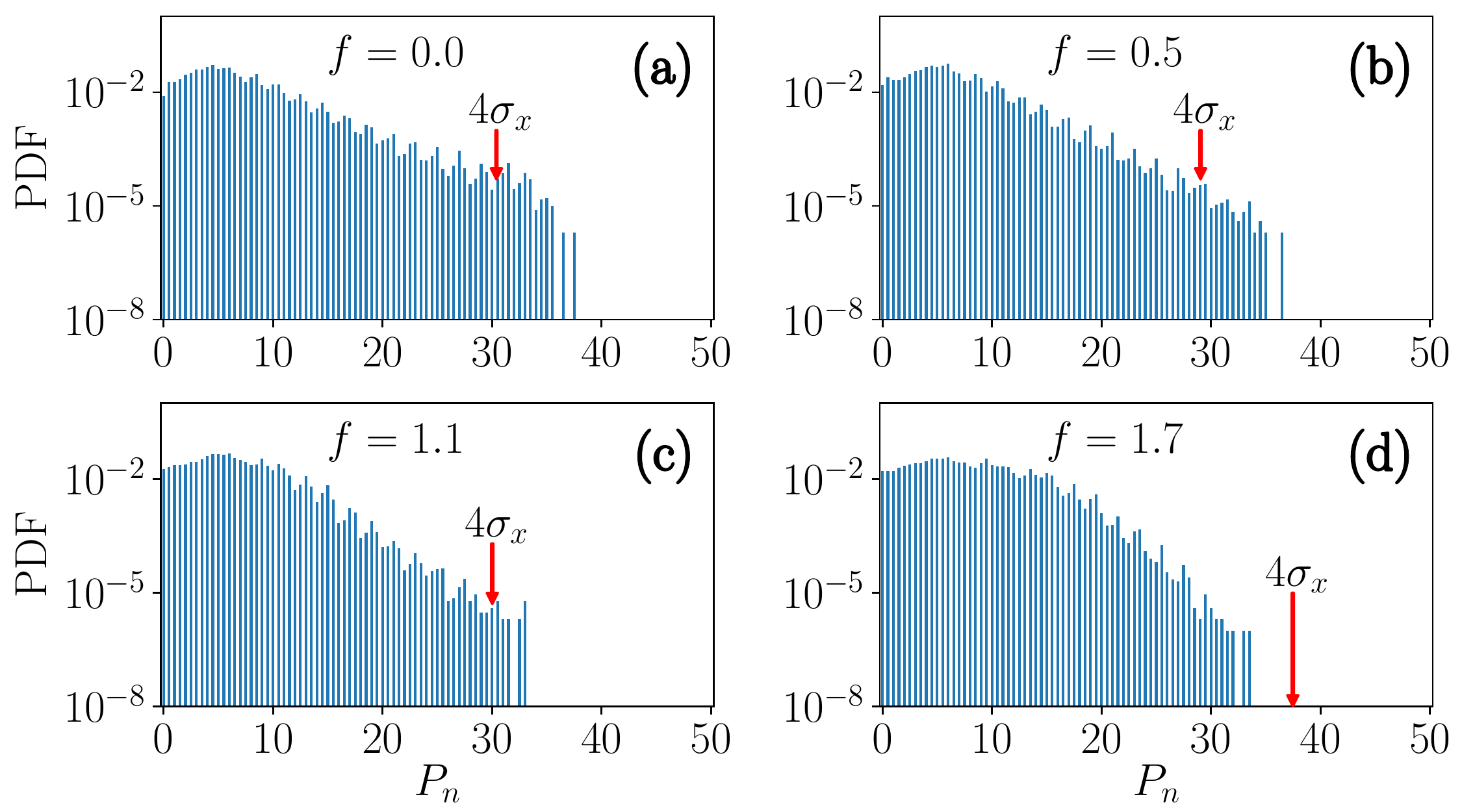}
	\caption{Plots of the PDF of peaks for different values of the external forcing amplitude $f$: (a) $f = 0$, (b) $f=0.05$, (c) $f=1.1$ and (d) $f=1.7$.. The other parameters are the same as in Fig.~\ref{sup}. The qualifier threshold $x_{ee}=\langle x \rangle + 4\sigma_x$ is noted by red vertical arrows and represented just as $4\sigma_x$.\color{black}}
	\label{pdffor}
\end{figure}%
Further, to have a clear picture, we plot the distribution (PDF) of the peaks ($P_n$) in Figs.~\ref{pdffor}(a) - \ref{pdffor}(d) for four different values of $f$, namely $f=0.0$, $0.5$, $1.1$ and $1.7$. The qualifier threshold $x_{ee}=\langle x \rangle + 4\sigma_x$ is noted by red vertical arrows and represented just as $4\sigma_x$. From Fig.~\ref{pdffor}, we can see the decreasing effect of extreme events more clearly while increasing the forcing strength. Figure~\ref{pdffor}(b) has comparatively lesser number of peaks crossing the threshold when compared to Fig.~\ref{pdffor}(a). Similarly, the number of peaks crossing the threshold in Fig.~\ref{pdffor}(c) is comparatively less when compared with Fig.~\ref{pdffor}(b). For $f=1.7$, no peaks are present beyond the threshold which is evident from Fig.~\ref{pdffor}(d).

\begin{figure}[!ht]
	\centering
	\includegraphics[width=0.5\textwidth]{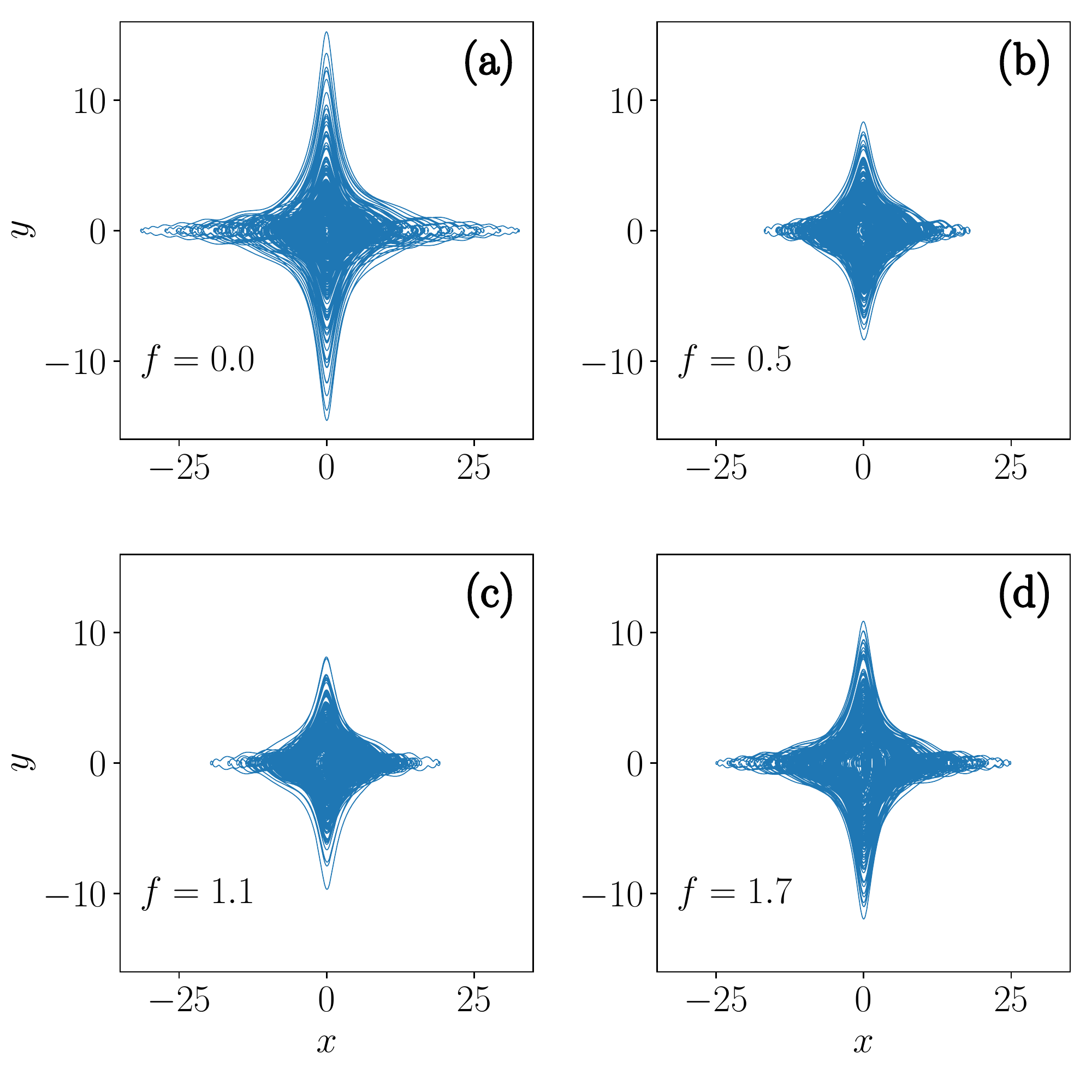}
	\caption{Plots of the phase portraits in $x - y$ plane for different values of the external forcing amplitude $f$: (a) $f = 0$, (b) $f=0.05$, (c) $f=1.1$ and (d) $f=1.7$. The other parameters are fixed as $\lambda=0.5$, $\omega_0^2=0.25$, $\Omega_0^2=6.7$, $\omega_p=1.0$, $\alpha=0.2$, $\epsilon = 0.081$ and $\omega_e = 1$.}
	\label{forphase}
\end{figure}

To understand the dynamical changes happening in system (\ref{pardri}) under the influence of external forcing, we plot the phase portrait of the system at four different values of the forcing strength, namely $f=0.0$, $0.5$, $1.1$ and $1.7$ in Figs.~\ref{forphase}(a) - \ref{forphase}(d). %
The value of $\epsilon$ is fixed at $0.081$. We can visualize from Fig.~\ref{forphase} that the size of phase portrait decreases with the introduction of external forcing. Further the presence of the single lengthy trajectories in the outer layer of the phase portrait (which is responsible for the production of extreme events, see Figs.  \ref{forphase}(a), \ref{forphase}(b) and \ref{forphase}(c)) decreases and when $\epsilon=1.7$ there are only dense trajectories (because of which there are no extreme events produced, see Fig. \ref{forphase}(d)). An important point to note here is that the shape of the phase portrait does not get altered even under the influence of the external forcing. It means, even with a larger amplitude of the external force, the velocity dependent potential's nature still dominates in the system. Thus, the suppression of extreme events can be attributed to the fact that the application of external forcing increases the velocity of the system, reducing the amplitude of $x$ thereby mitigating the occurrence of extreme events.

\begin{figure}[!ht]
	\centering
	\includegraphics[width=0.5\textwidth]{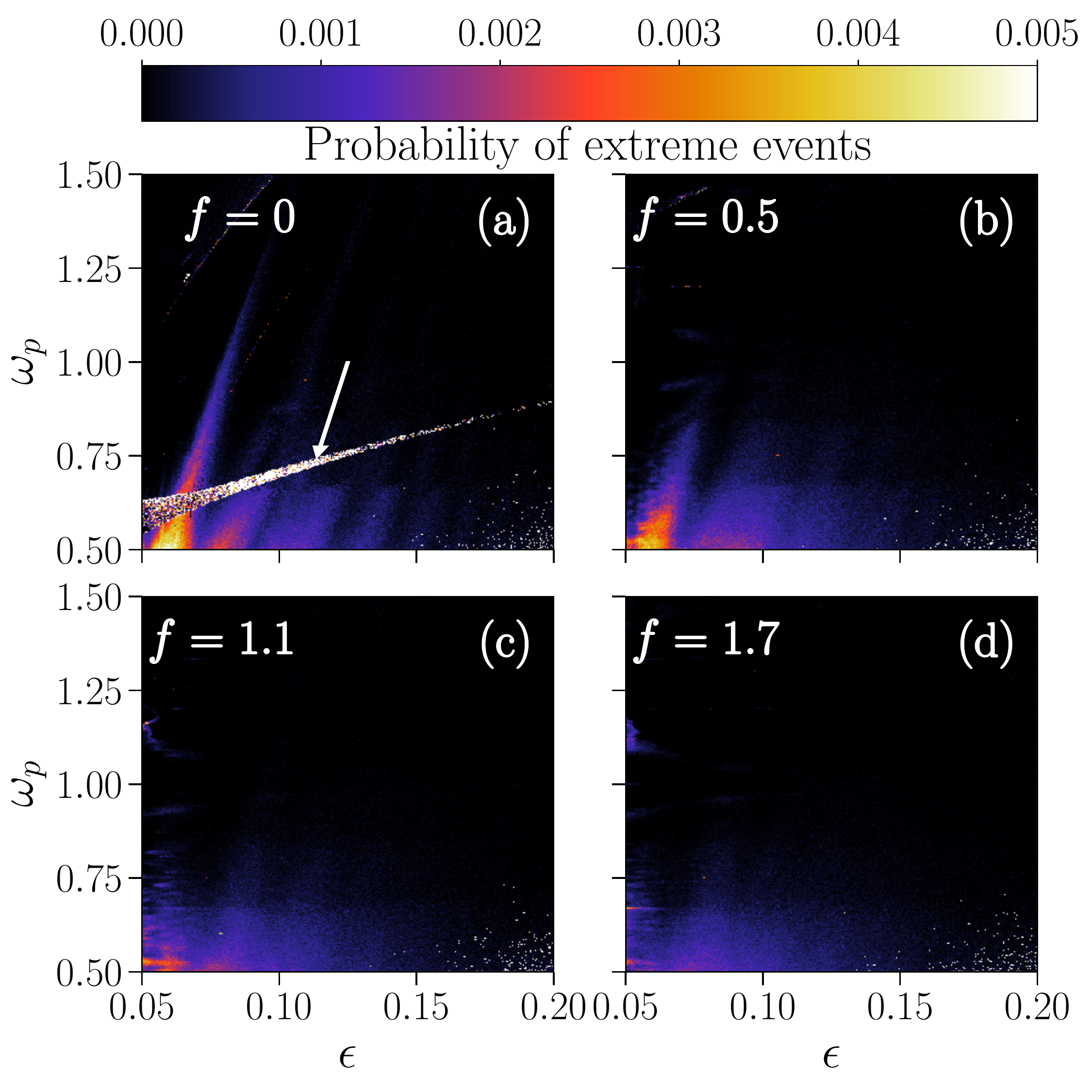}
	\caption{Two parameter diagrams in $\epsilon - \omega_p$ plane for various values of the external forcing strength: (a) $f = 0$, (b) $f=0.05$, (c) $f=1.1$ and (d) $f=1.7$. 
		The other parameters are fixed as $\lambda=0.5$, $\omega_0^2=0.25$, $\Omega_0^2=6.7$, $\omega_p=1.0$, $\alpha=0.2$, and $\omega_e = 1$. The colour bar in Fig. 13 represents the variation in the probability of extreme events where black colour represents regions with zero extreme events, and the white colour represents regions with a probability of $5 \times 10^{-3}$ and above. 
	}
	\label{2phase}
\end{figure}

Figure~\ref{2phase}(a) corresponds to a two parameter diagram of the probability in the $\epsilon - \omega_p$ plane showing the regions of the occurrence of extreme events in the absence of external force $f=0$. The complete picture of the system being influenced by three different external forcing strengths $f=0.05$, $f=1.1$ and $f=1.7$ is given in the two parameter diagrams shown in Figs.~\ref{2phase}(a) - \ref{2phase}(d) corresponding to the values $f=0$, $0.5$, $1.1$ and $1.7$, respectively. We have extended our analysis in the two parameter diagram up to $\epsilon=0.20$. %
The colour bar in Fig. 13 represents the variation in the probability of extreme events where black colour represents regions with zero extreme events, and the white colour represents regions with a probability of $5 \times 10^{-3}$ and above. An important property which we point out here is that extreme events alternatively increase and decrease alongside $\omega_p=1.0$ (previously discussed in Sec. \ref{sec:4}), can be seen from the violet (grey) cloud like structure formed alternatively. This cloud like structure forms as an extension of the tongue like structure starting around $\omega_p=0.5$, and $\epsilon=0.05$. Although it is broader at the base, it sharpens while moving towards top right in the parameter region. Further, it is easily visible from Fig.~\ref{2phase}(a) that regions of zero extreme events are placed alternatively between these tongs. A notable property is that extreme events are suppressed in (\ref{pardri}) when it is subjected to an external force, which is visible in the two parameter diagram Figs.~\ref{2phase}(b) - \ref{2phase}(d). This can be attributed from the systematic destruction of the tongs while increasing the value of the external forcing strength. Especially when $f=1.7$ extreme events are suppressed in many regions and the tong like structures are almost destroyed. This can be seen from Fig.~\ref{2phase}(d). In particular, almost all regions above $\omega_p=1.0$ that displayed the extreme events previously are now suppressed. Although a few regions show extreme events (around $\omega_p=1.2$ and $\epsilon=0.05$) by the influence of external forcing, it is comparatively less than the regions where extreme events are suppressed. The scattered white points near the right bottom ($(\epsilon, \omega_p) = (0.2, 0.5)$) of Fig. {\ref{2phase}} represent the regions displaying a large number of extreme events with very high probabilities. This can be seen for all the four values of $f$. One more point to note here is that the white shaded tong-like structure basing around $\epsilon=0.05$, $\omega_p=0.6$ and with a tip near $\epsilon=0.2$, $\omega_p=0.9$, indicated by a white arrow in Fig.~\ref{2phase}(a), does not correspond to the extreme events. Rather it represents the regions of long transient behaviour similar to what we had seen in Figs. \ref{fig6} and \ref{sup}. In these regions, the system either transits to periodic orbits or approaches to fixed points after a long chaotic transient. So while we engineer systems using this model, the parameter regions can be chosen aptly to avoid extreme events.

\begin{figure}[!ht]
	\centering
	\includegraphics[width=0.5\textwidth]{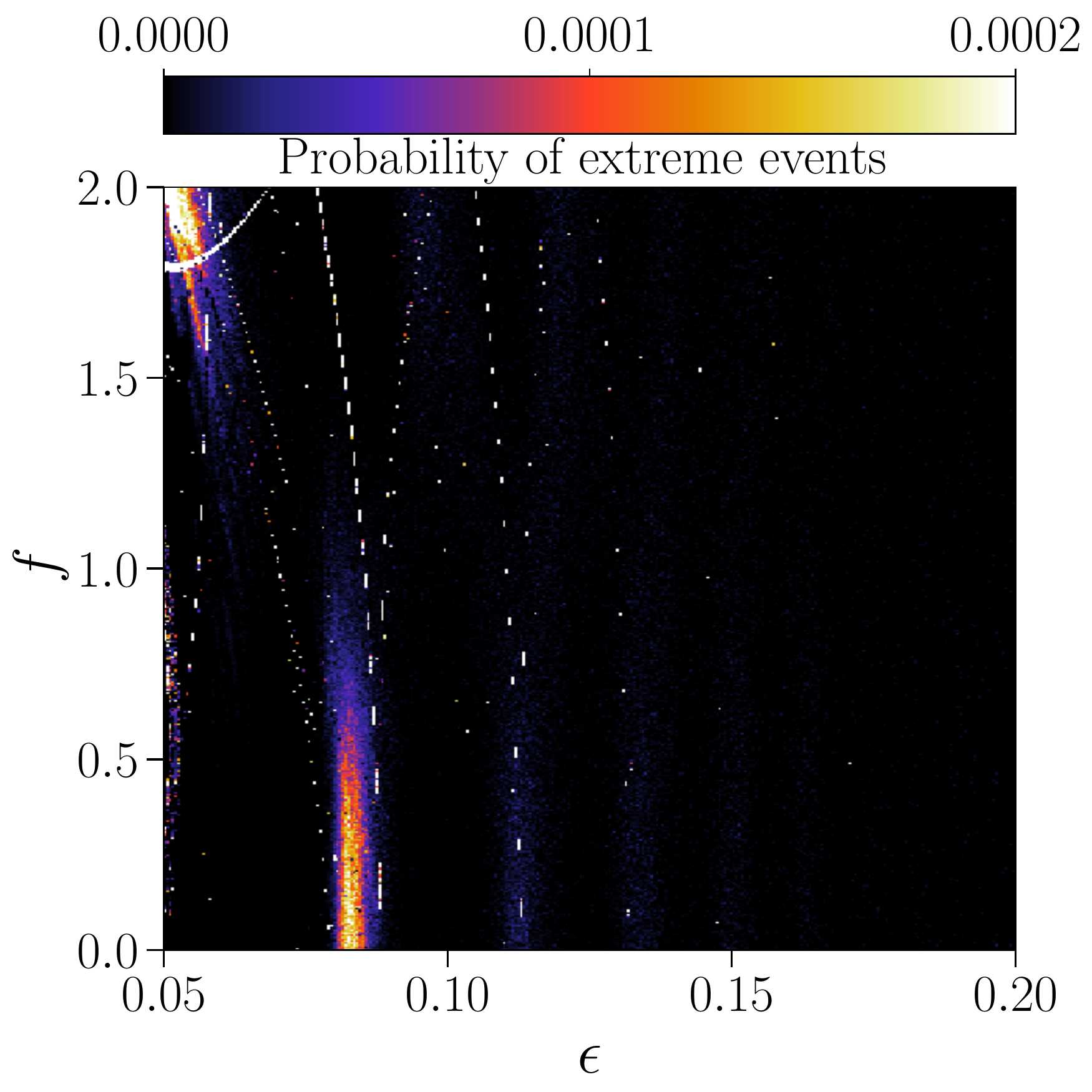}
	\caption{Two parameter diagrams in $\epsilon - f$ plane. The other parameters are fixed as $\lambda=0.5$, $\omega_0^2=0.25$, $\Omega_0^2=6.7$, $\omega_p=1.0$, $\alpha=0.2$, and $\omega_e = 1$. The colour bar in Fig. \ref{ef} represents the variation in the probability of extreme events where black colour represents regions with zero extreme events, and the white colour represents regions with a probability of $2 \times 10^{-4}$ and above.
	}
	\label{ef}
\end{figure}%

Figure (\ref{ef}) represent the two parameter diagram in the $(\epsilon-f)$ plane. The colour bar in Fig. \ref{ef} represents the variation in the probability of extreme events where black colour represents regions with zero extreme events, and the white colour represents regions with a probability of $2 \times 10^{-4}$ and above. From the figure, we can see the diminishing effect of extreme events at $\epsilon=0.081$ as $f$ increases (similar to the one we observed in Fig.~\ref{sup}). Regions where the probability of extreme events alternates with a brief region displaying zero probability (as discussed in Sec. \ref{sec:4}) is visible as alternating violet shades (dark grey) in Figure (\ref{ef}). Suppression of extreme events occur in all the alternating regions. In the first alternating region around $\epsilon=0.081$, we can observe the suppresion of extreme events from high probability. This is why we can observe the transition form orange (light grey) to violet (dark grey) to black. Whereas in all the other alternating regions, the probability is comparatively less and the suppression occurs earlier as shown in the transition from violet (dark grey) to black. Hence, the suppressing nature of the external force, particularly in system (\ref{pardri}), can be clearly seen in Fig. (\ref{ef}). 

\color{black}

Thus, extreme events that are present in a parametrically driven nonlinear non-polynomial oscillator with velocity dependent potential can be suppressed by the addition of external forcing.

\section{Effect of noise}
\label{sec:noise}

Noise can both induce and suppress extreme events depending on the system and parameters under consideration \cite{ZamoraMunt2014,ZamoraMunt2013,ZamoraMunt2014a,Ahuja2014}. Here, we study the effect of noise in the system (\ref{pardri}) under two settings $(i)$ when noise is added to the state variable $x$ as in  Ref.~\cite{Rajasekar2016} and $(ii)$ when noise is added directly to the equation (\ref{pardri}) as in Ref.~\cite{ZamoraMunt2013}. In case $(i)$ the noise is included as $x(t+\Delta t)~\rightarrow~x(t+\Delta t)+\sqrt{D\Delta t}~\xi(t)$, where $\xi(t)$ represents Gaussian random numbers with zero mean and variance $D$, where $D$ represents the noise strength. In case $(ii)$ also, Gaussian noise is considered. In case $(i)$, extreme events become super-extreme \cite{Bonatto2017} when the strength of noise increases. Here we declare extreme events to be super-extreme since it qualifies even the $\mu+16\sigma_x$ threshold. 
\begin{figure}[!ht]
	\centering
	\includegraphics[width=0.5\textwidth]{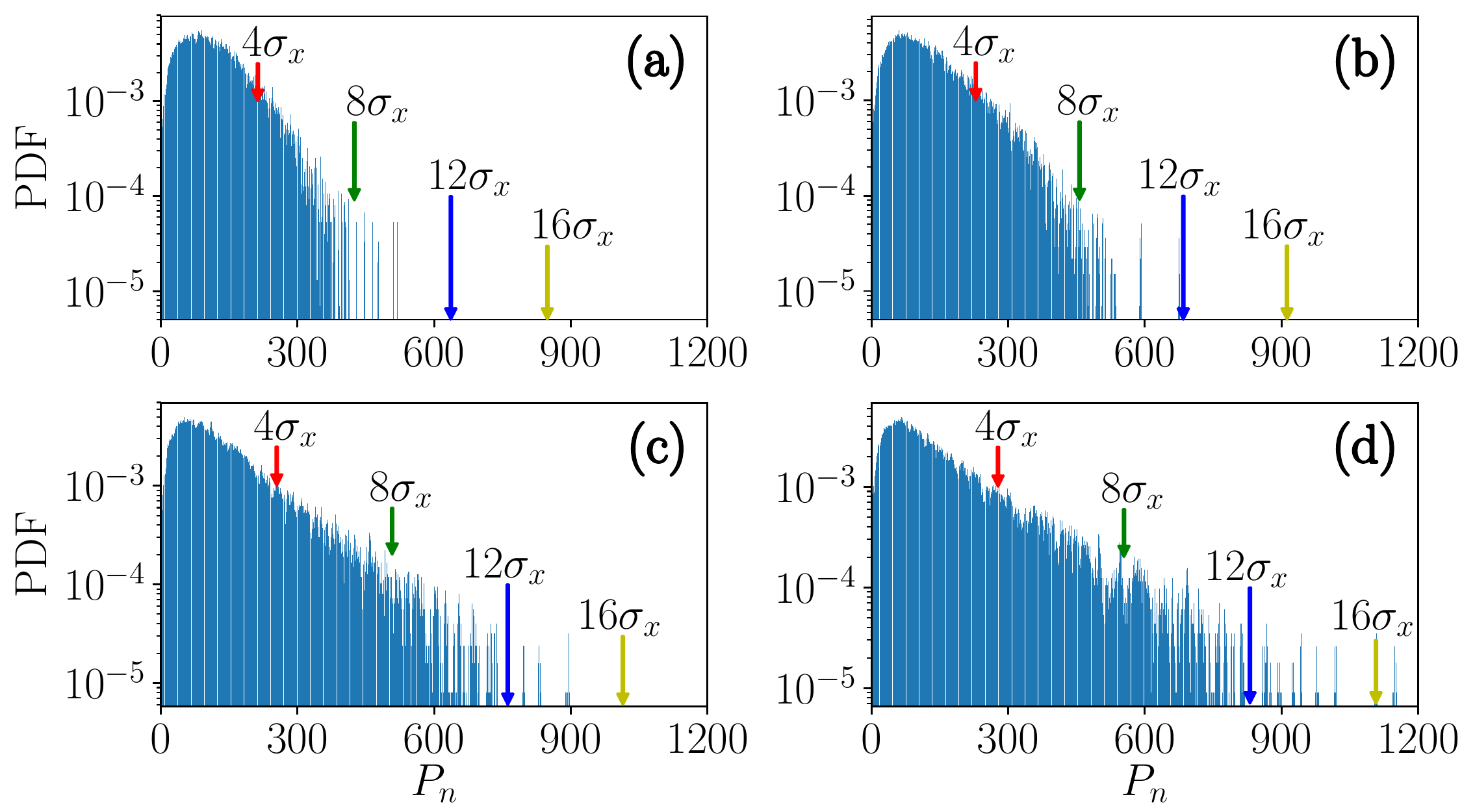}
	\caption{Plot showing the PDF of peaks $(P_n)$ for different noise strengths: (a) $D=0.0$, (b) $D=0.05$, (c) $D=0.2$, and (d) $D=0.3$. The other parameters are $\epsilon=0.2$, $F=0.0$, $\omega_p=0.5$. The arrows denote the four threshold qualifiers, namely  $4\sigma_x$, $8\sigma_x$, $12\sigma_x$ and $16\sigma_x$.}
	\label{noise}
\end{figure}%
Figures \ref{noise}(a) - \ref{noise}(d) illustrate the effect of noise on the probability of the occurrence of extreme events. The arrows in Fig.~\ref{noise} marks the $4\sigma_x$, $8\sigma_x$, $12\sigma_x$ and $16\sigma_x$ threshold qualifiers. Although, we have made our entire analysis only with the $4\sigma_x$ qualifier threshold, we also identify that in the regions of the red-scattered points in Fig. \ref{2phase}, it exhibits extreme events for $8\sigma_x$ threshold, as shown in Fig. \ref{noise}(a). Then as the strength of noise increases, we can see the emergence of super-extreme events similar to the event observed earlier in a parametrically driven loss modulated CO$_2$ laser in the absence of noise \cite{Bonatto2017}. In Fig. \ref{noise}(b) extreme events can occur only up to  a threshold qualifier value of $8\sigma_x$ when $D=0.05$. On increasing the noise strength to $D=0.2$, extreme events even qualify the $12\sigma_x$ threshold qualifier mark, as shown in Fig.~\ref{noise}(c). Further, when $D=0.3$, extreme events cross the $16\sigma_x$ threshold as shown in Fig.~\ref{noise}(d). As mentioned before, this occurs at the regions of scattered red points shown in Fig. \ref{2phase}. For the case $(ii)$, when the noise is added to the equation, we witness noise-induced extreme events but the probability remains almost constant when increasing the noise strength and no super-extreme events are observed. In both the cases, the addition of external forcing does not produce any noticeable change in the probability of the occurrence of extreme events. Thus, in our case, the introduction of noise is highly detrimental, especially when introduced to the state variable. Eventhough noise supresses extreme events in many systems, in our case extreme events persists under noise and even transforms to super-extreme events.

\section{Conclusion}
\label{sec:6}
In this work, to begin, we have investigated the emergence of extreme events in a parametrically driven nonlinear non-polynomial mechanical model with velocity-dependent potential without external forcing. We have confirmed the emergence of extreme events using the peak probability distribution plot, which shows a fat-tail distribution, a key signature for the occurrence of extreme events. The extreme event does not occur in the regions of interior crisis, but it occurs in the places where the chaotic attractor expands continuously. We identified a viable mechanism for the occurrence of extreme events in this model which is mainly due to the change in velocity of the system. In other words extreme events occur whenever the velocity of the system is minimum. The chaotic attractor alternatively expands and contracts which is due to the alternate decrease and increase in the velocity of the system. In those corresponding regions, the probability of the extreme events also increases and decreases alternatively with a brief neighbourhood of zero-probability as the strength of the parametric drive is varied. This fact is also confirmed using the peak distribution and the probability plot. While subjecting the system (\ref{pardri}) to an external drive, we found that the extreme events get mitigated. That is the probability of the occurrence of extreme events decreases to zero at a particular point while varying the external forcing strength. This result is the most intriguing because driving a system with an external force is the simplest perturbation that one can incorporate in any physical, electronic and mechanical systems and surprisingly it suppresses the extreme event. This mitigation happened due to the increase in the velocity of the system under external forcing.  Thus, the velocity plays a pivotal role in the production and suppression of extreme events in the systems (\ref{pardri}). We have also observed the occurrence of super-extreme events under the influence of noise. From our study, we also conclude that the presence of noise does not have any suppression effect on the extreme event, rather it turns the extreme events into super-extreme events in a particular case. This work may be a starting point in considering external force as an useful tool to control extreme events.
\color{black}
\section*{Acknowledgement}
SS  thanks  the  Department  of  Science and  Technology (DST), Government of India, for support through INSPIRE Fellowship (IF170319).  The work of AV forms a part of a research project sponsored by DST under the Grant No. EMR/2017/002813. The work of PM forms parts of sponsored research projects by Council of Scientific and Industrial Research (CSIR), India (Grant No. 03(1422)/18/EMR-II), and Science and Engineering Research Board (SERB), India (Grant No. CRG/2019/004059). The work of MS forms a part of a research project sponsored by CSIR, India under the Grant No. 03(1397)/17/EMR-II.


\begin{thebibliography}{50}
	
	\bibitem{Krause2015}	
	S.M. Krause, S. B\"orries, S. Bornholdt, Phys. Rev. E \textbf{92}, 012815 (2015).
	
	\bibitem{Dysthe2008}
	K.  Dysthe,  H.  E.  Krogstad, P.  M\"uller, Annu.  Rev.  Fluid Mech. \textbf{40}, 287 (2008).
	
	\bibitem{Jentsch2005} 
	S.A.V. Jentsch, H. Kantz (eds.),Extreme Events in Nature and Society (Springer, Heidelberg, 2005).
	
	\bibitem{Kumarasamy2018}
	S.~Kumarasamy, A.N. Pisarchik, Phys. Rev. E \textbf{98}, 032203 (2018).
	
	\bibitem{Ansmann2013}
	G.~Ansmann, R.~Karnatak, K.~Lehnertz, U.~Feudel, Phys. Rev. E \textbf{88},
	052911 (2013).
	
	\bibitem{Reinoso2013}
	J.A. Reinoso, J.~Zamora-Munt, C.~Masoller, Phys. Rev. E \textbf{87}, 062913
	(2013).
	
	\bibitem{Solli2007}
	D.R. Solli, C.~Ropers, P.~Koonath, B.~Jalali, Nature (London) \textbf{450},
	1054 (2007).
	
	\bibitem{Bodai2011}
	T.~B{\'{o}}dai, G.~K{\'{a}}rolyi, T.~T{\'{e}}l, Nonlin. Processes Geophys.
	\textbf{18}, 573 (2011).
	
	\bibitem{Yukalov2012}
	V.I. Yukalov, E.P. Yukalova, D.~Sornette, Eur. Phys. J. Spec. Top.
	\textbf{205}, 313 (2012).
	
	\bibitem{Moitra2019}
	P.~Moitra, S.~Sinha, Chaos \textbf{29}, 023131 (2019).
	
	\bibitem{Chaurasia2020}
	S.S. Chaurasia, U.K. Verma, S.~Sinha, Sci. Rep. \textbf{10}, 10613 (2020).
	
	\bibitem{Kingston2017}
	S.L. Kingston, K.~Thamilmaran, P.~Pal, U.~Feudel, S.K. Dana, Phys. Rev. E
	\textbf{96}, 052204 (2017).
	
	\bibitem{Kingston2020}
	S.L. Kingston, K.~Suresh, K.~Thamilmaran, T.~Kapitaniak, Eur. Phys. J. Spec.
	Top. \textbf{229}, 1033 (2020).
	
	\bibitem{Ray2020}
	A.~Ray, A.~Mishra, D.~Ghosh, T.~Kapitaniak, S.K. Dana, C.~Hens, Phys. Rev. E
	\textbf{101}, 032209 (2020).
	
	\bibitem{Chowdhury2019}
	S.N. Chowdhury, S.~Majhi, M.~Ozer, D.~Ghosh, M.~Perc, New J. Phys, \textbf{21},
	073048 (2019).
	
	\bibitem{Karnatak2014}
	R.~Karnatak, G.~Ansmann, U.~Feudel, K.~Lehnertz, Phys. Rev. E \textbf{90},
	022917 (2014).
	
	\bibitem{Saha2017}
	A.~Saha, U.~Feudel, Phys. Rev. E \textbf{95}, 062219 (2017).
	
	\bibitem{Saha2018}
	A.~Saha, U.~Feudel, Chaos \textbf{28}, 033610 (2018).
	
	\bibitem{Rings2017}
	T.~Rings, G.~Ansmann, K.~Lehnertz, Eur. Phys. J. Spec. Top. \textbf{226}, 1963
	(2017).
	
	\bibitem{Bialonski2015}
	S.~Bialonski, G.~Ansmann, H.~Kantz, Phys. Rev. E \textbf{92}, 042910 (2015).
	
	\bibitem{Ansmann2016}
	G.~Ansmann, K.~Lehnertz, U.~Feudel, Phys. Rev. X \textbf{6}, 011030 (2016).
	
	\bibitem{Mishra2018}
	A.~Mishra, S.~Saha, M.~Vigneshwaran, P.~Pal, T.~Kapitaniak, S.K. Dana, Phys.
	Rev. E \textbf{97}, 062311 (2018).
	
	\bibitem{Oliveira2016}
	G.F. de~Oliveira, O.~Di~Lorenzo, T.P. de~Silans, M.~Chevrollier, M.~Ori\'a,
	H.L.D. de Souza Cavalcante, Phys. Rev. E \textbf{93}, 062209 (2016).
	
	\bibitem{Ray2019}
	A.~Ray, S.~Rakshit, D.~Ghosh, S.K. Dana, Chaos \textbf{29}, 043131 (2019).
	
	\bibitem{Cousins2014}
	W.~Cousins, T.P. Sapsis, Physica D \textbf{280-281}, 48 (2014).
	
	\bibitem{Kim2003}
	J.W. Kim, E.~Ott, Phys. Rev. E \textbf{67}, 026203 (2003).
	
	\bibitem{Galuzio2014}
	P.P. Galuzio, R.L. Viana, S.R. Lopes, Phys. Rev. E \textbf{89}, 040901 (2014).
	
	\bibitem{Bailung2011}
	H.~Bailung, S.K. Sharma, Y.~Nakamura, Phys. Rev. Lett. \textbf{107}, 255005
	(2011).
	
	\bibitem{Ganshin2008}
	A.N. Ganshin, V.B. Efimov, G.V. Kolmakov, L.P. Mezhov-Deglin, P.V.E.
	McClintock, Phys. Rev. Lett. \textbf{101}, 065303 (2008).
	
	\bibitem{Pisarchik2018}
	A.N. Pisarchik, V.V. Grubov, V.A. Maksimenko, A.~Lüttjohann, N.S. Frolov,
	C.~Marqu{\'{e}}s-Pascual, D.~Gonzalez-Nieto, M.V. Khramova, A.E. Hramov, Eur.
	Phys. J. Spec. Top. \textbf{227}, 921 (2018).
	
	\bibitem{Toffoli2017}
	A.~Toffoli, D.~Proment, H.~Salman, J.~Monbaliu, F.~Frascoli, M.~Dafilis,
	E.~Stramignoni, R.~Forza, M.~Manfrin, M.~Onorato, Phys. Rev. Lett.
	\textbf{118}, 144503 (2017).
	
	\bibitem{Sapsis2018}
	T.P. Sapsis, Phil. Trans. R. Soc. A \textbf{376}, 20170133 (2018).
	
	
	\bibitem{S.Cavalcante2013}
	H.L.D. de Souza Cavalcante, M.~Ori\'a, D.~Sornette, E.~Ott, D.J. Gauthier, Phys.
	Rev. Lett. \textbf{111}, 198701 (2013).
	
	\bibitem{Pisarchik2011}
	A.N. Pisarchik, R.~Jaimes-Re\'ategui, R.~Sevilla-Escoboza, G.~Huerta-Cuellar,
	M.~Taki, Phys. Rev. Lett. \textbf{107}, 274101 (2011).
	
	
	\bibitem{ZamoraMunt2014}
	J.~Zamora-Munt, C.R. Mirasso, R.~Toral, Phys. Rev. E \textbf{89}, 012921 (2014).
	
	\bibitem{ZamoraMunt2013}
	J.~Zamora-Munt, B.~Garbin, S.~Barland, M.~Giudici, J.R.R. Leite, C.~Masoller,
	J.R. Tredicce, Phys. Rev. A \textbf{87}, 035802 (2013).
	
	\bibitem{ZamoraMunt2014a}
	J.~Zamora-Munt, S.~Perrone, R.~Vilaseca, C.~Masoller, Advanced Photonics JM5A.51 (2014).
	
	\bibitem{Ahuja2014}
	J.~Ahuja, D.B. Nalawade, J.~Zamora-Munt, R.~Vilaseca, C.~Masoller, Opt. Express
	\textbf{22}, 28377 (2014).
	
	\bibitem{Suresh2018}
	R.~Suresh, V.K. Chandrasekar, Phys. Rev. E \textbf{98}, 052211 (2018).
	
	\bibitem{Chen2014}
	Y.Z. Chen, Z.G. Huang, Y.C. Lai, Sci. Rep. \textbf{4}, 6121 (2014).
	
	\bibitem{Farazmand2019}
	M.~Farazmand, T.P. Sapsis, Phys. Rev. E \textbf{100}, 033110 (2019).
	
	\bibitem{Kaveh2020}
	H.~Kaveh, H.~Salarieh, Chaos, Solitons Fractals \textbf{136}, 109827 (2020).
	
	\bibitem{Nayfeh1979}
	A.H. Nayfeh, D.T. Mook, \emph{Nonlinear oscillations} (Wiley, New York, 1979).
	
	\bibitem{Venkatesan1997}
	A.~Venkatesan, M.~Lakshmanan, Phys. Rev. E \textbf{55}, 5134 (1997).
	
	\bibitem{Venkatesan1998}
	A.~Venkatesan, M.~Lakshmanan, Phys. Rev. E \textbf{58}, 3008 (1998).
	
	\bibitem{Sanderson1977}
	R.~Sanderson, K.E. Bird, Methods in Cell Biology \textbf{15}, 1  (1977).
	
	\bibitem{Bear1984}
	J.~Bear, M.Y. Corapcioglu, J.~Balakrishna, Adv. Water Resources \textbf{7}, 150
	(1984).
	
	\bibitem{Lai1997}
	P.Y. Lai, L.C. Jia, C.K. Chan, Phys. Rev. Lett. \textbf{79}, 4994 (1997).
	
	\bibitem{Bonatto2017}
	C.~Bonatto, A.~Endler, Phys. Rev. E \textbf{96}, 012216 (2017).
	
	\bibitem{Metayer2014}
	C.~Metayer, A.~Serres, E.J. Rosero, W.A.S. Barbosa, F.M. de~Aguiar, J.R.R.
	Leite, J.R. Tredicce, Opt. Express \textbf{22}, 19850 (2014).
	
	\bibitem{DELBOURGO1969}
	R.~Delbourgo, A.~Salam, J.~Strathdee, Phys. Rev. \textbf{187}, 1999 (1969).
	
	\bibitem{Wolf1985}
	A. Wolf, J.B. Swift, H.L. Swinney, J.A. Vastano, Physica D \textbf{16}, 285 (1985).
	
	\bibitem{Rajasekar2016}
	S.~Rajasekar, M.~Sanju\'{a}n, \emph{Nonlinear Resonances}, Springer Series in
	Synergetics (Springer International Publishing, 2016)
	
\end{thebibliography}
\end{document}